\def\simlt{\mathrel{\spose{\lower 3pt\hbox{$\mathchar''218$}}
     \raise 2.0pt\hbox{$\mathchar''13C$}}}
\def\simgt{\mathrel{\spose{\lower 3pt\hbox{$\mathchar''218$}}
     \raise 2.0pt\hbox{$\mathchar''13E$}}}
\documentstyle[aaspp4,epsf]{article} 
\begin{document}
\def\gtorder{\mathrel{\raise.3ex\hbox{$>$}\mkern-14mu
             \lower0.6ex\hbox{$\sim$}}}
\def\ltorder{\mathrel{\raise.3ex\hbox{$<$}\mkern-14mu
             \lower0.6ex\hbox{$\sim$}}}

\def\today{\number\year\space \ifcase\month\or  January\or February\or
        March\or April\or May\or June\or July\or August\or
        September\or
        October\or November\or December\fi\space \number\day}
\def\fraction#1/#2{\leavevmode\kern.1em
 \raise.5ex\hbox{\the\scriptfont0 #1}\kern-.1em
 /\kern-.15em\lower.25ex\hbox{\the\scriptfont0 #2}}
\def\spose#1{\hbox to 0pt{#1\hss}}
\def\rsun{$R_{\odot}$}
\def\msun{$M_{\odot}$}
\def\rstar{$R_{star}$}
\def\rplanet{$R_{p}$}
\title{The EXPLORE Project I:  A Deep Search for Transiting Extrasolar Planets}

\author{G.\ Mall\'en-Ornelas\footnote{Princeton University Observatory, 
Peyton Hall, Princeton, NJ 08544;
mallen@astro.princeton.edu}$^{,}$\footnote{Departamento de
Astronom\'{\i}a y Astrof\'{\i}sica, Pontificia Universidad Cat\'olica
de Chile, Casilla 306, Santiago 22, Chile}$^{,}$\footnote{Visiting
Astronomer, Cerro Tololo Inter-American Observatory, National Optical
Astronomy Observatories, which are operated by the Association of
Universities for Research in Astronomy, under contract with the
National Science Foundation},
S.\ Seager\footnote{Institute for Advanced Study, Einstein
Drive, Princeton, NJ 08540; present address: The Carnegie 
Institution of Washington, Dept. of Terrestrial Magnetism, 
5241 Broad Branch Rd. NW, Washington, DC 20015;
seager@dtm.ciw.edu},
H.\ K.\ C.\ Yee$^{3,}$\footnote{Department of Astronomy and Astrophysics, University of Toronto, 60
St.\ George St., Toronto, ON M5S 3H8, Canada; hyee@astro.utoronto.ca},
D.\ Minniti$^{3,}$\footnote{Departamento de Astronom\'{\i}a y
Astrof\'{\i}sica, Pontificia Universidad Cat\'olica de Chile, Casilla
306, Santiago 22, Chile; dante@astro.puc.cl},
Michael D.\ Gladders$^{5,}\footnote{Present address: The Carnegie Observatories, 813
Santa Barbara St., Pasadena, CA 91107; gladders@ociw.edu}$,
G.\ M.\ Mall\'en-Fullerton\footnote{Universidad Iberoamericana, 
Prolongaci\'on Paseo de la Reforma 880, Edificio F Segundo Piso, Col.
Lomas de Santa Fe, 01200 M\'exico, D.F., M\'exico;
guillermo.mallen@uia.mx}, T.\ M.\ Brown\footnote{High Altitude
Observatory/National Center for Atmospheric Research, P.O.\ Box 3000,
Boulder, CO 80307; timbrown@hao.ucar.edu}}

\pagestyle{plain}

\begin{abstract}

Planet transit searches promise to be the next breakthrough for
extrasolar planet detection, and will bring the characterization of
short-period planets into a new era.  Every transiting planet
discovered will have a measured radius, which will provide constraints
on planet composition, evolution, and migration history.  Together
with radial velocity measurements, the absolute mass of every
transiting planet will be determined.

In this paper we discuss the design considerations of the EXPLORE
(EXtrasolar PLanet Occultation REsearch) project, a series of
transiting planet searches using 4-m-class telescopes to continuously
monitor a single field of stars in the Galactic Plane in each $\sim2$
week observing campaign.  We discuss the general factors which
determine the efficiency and the number of planets found by a transit
search, including time sampling strategy and field selection.  The
primary goal is to select the most promising planet candidates for
radial velocity follow-up observations.  We show that with very high
photometric precision light curves that have frequent time sampling
and at least two detected transits, it is possible to uniquely solve
for the main parameters of the eclipsing system (including planet
radius) based on several important assumptions about the central star.
Together with a measured spectral type for the star, this unique solution for
orbital parameters provides a powerful method for ruling out most
contaminants to transiting planet candidates.  For the EXPLORE
project, radial velocity follow-up observations for companion mass
determination of the best candidates are done on 8-m-class telescopes
within two or three months of the photometric campaigns.  This
same-season follow-up is made possible by the use of efficient
pipelines to produce high quality light curves within weeks of the
observations.  We conclude by presenting early results from our first
search, EXPLORE I, in which we reached $<1$\% rms photometric precision
(measured over a full night) on $\sim$37,000 stars with $14.5 \le I \le 18.2$.
\end{abstract}

\section{Introduction}
The discovery of giant extrasolar planets in the mid-1990s using
radial velocity techniques (e.g., Marcy, Cochran, \& Mayor 2000)
heralded a new era in the study of planetary systems, and to date
$\sim$80 extrasolar giant planets have been
discovered\footnote{Extrasolar Planets Encyclopaedia,
http://www.obspm.fr/encycl/catalog.html}.  Radial velocity searches
produced the completely unexpected discovery of massive planets in
few-day period orbits, such as 51 Peg b (Mayor \& Queloz 1995).  To
date 17 systems with orbital distances of $<$0.1 AU and periods
of a few days have been found$^{10}$.  The existence of a class of
close-in giant planets shows that planetary systems can be radically
different from our own. The discovery of close-in giant planets
sparked much theoretical work on planet formation and migration
scenarios to explain the proximity of giant planets to the parent
star, such as: planetesimal scattering (e.g., Murray et al.\ 1998),
planet-disk or binary star interactions (e.g., Lin, Bodenheimer, \&
Richardson 1996; Holman, Touma, \& Tremaine 1997), and dynamical
instabilities in multiple giant planet systems (Rasio \& Ford 1996).

\begin{figure}
\plotone{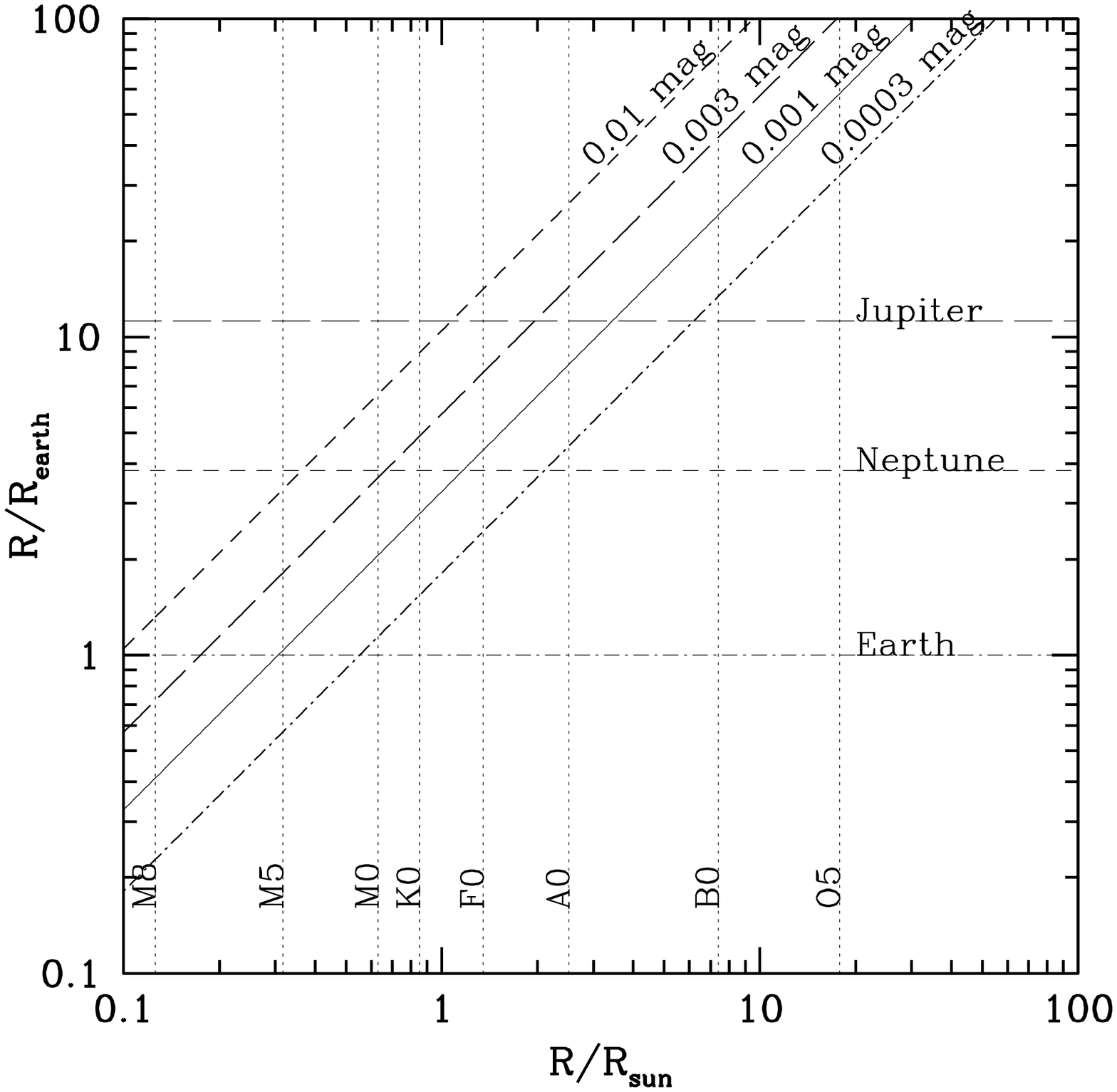}
\caption{Photometric precision requirements for detecting transits 
of planets with different radii, as a function of stellar radius.  The
diagonal lines indicate the combination of planet and star radii ($y$
and $x$ axes, respectively) which will result in transit depths
($R_p^2/R_*^2$) of 0.01, 0.003, 0.001, and 0.0003 mag.  If $N$ data
points are available during transit, a photometric precision equal to
the transit depth will result in a $N^{1/2} \sigma$ transit detection.
For example, a 1\% rms photometric precision will generally be
sufficient to detect all planet/star radius combinations that fall
above the top diagonal line.}
\label{fig:paramspace}
\end{figure}

The existence of a significant population of close-in extrasolar giant
planets (CEGPs) makes the method of finding planetary systems via
transits of their parent star very promising; the probability that a
given planet will show transits is inversely proportional to its
orbital distance, and relatively large for CEGPs around main-sequence
stars ($\sim$10\%).  Moreover, for planets with periods of 3--4 days,
it is possible to detect two or more transits via high photometric
precision light curves that span a relatively small number of nights.
A photometric precision of 1\%, which can be routinely achieved with
CCD cameras, is sufficient to detect giant planets around sun-like
stars (see Figure~\ref{fig:paramspace}).  The advent of wide-field CCD
mosaic cameras greatly increases the efficiency of the transit search
method, since a very large number of stars can be monitored at once.

The principal motivation for the transit search method is the
possibility of characterizing planets in a way not possible with
current radial velocity surveys: radius.  Transiting planets are
currently the only ones whose radii can be determined (based on
transit depth and stellar radius).  The planet radius can be measured
to better than 10\% precision with follow-up ultra-high precision
photomtery on transiting planets (e.g., Brown et al.\ 2001; Cody
\& Sasselov 2002).  A radius measurement is necessary to constrain
the planet evolution and migration history and also provides
constraints on planet composition and atmosphere through evolutionary
models. The radii of CEGPs are especially interesting for planetary
physics due to the evolution of the planet in proximity to the central
star and because of the as yet unknown migration timescales.  Very
interesting recent work (Guillot \& Showman 2002) has shown that the
observed radius of transiting planet HD209458b (Brown et al.\ 2001)
disagrees with the theoretically predicted radius by 30\% for their
preferred evolutionary models.  This implies there are atmospheric or
interior physical processes taking place which are not currently
known.  The determination of planet radius as a function of stellar
type and in different stellar environments will be a major step
forward in planetary characterization and in understanding giant
planet physics.


Follow-up observations of planets found from transit surveys can
be very important for planet confirmation and characterization.
Radial-velocity planet mass measurements are required in order to
confirm transit candidates as actual planets.  Mass measurements
for transiting planets are facilitated by the fact that the phase and
period for the system are known in advance, and therefore it is
possible to conduct observations at times of maximum radial velocity
amplitude.  For faint stars (V$\sim 18$) it is currently possible to
measure masses or place upper limits as low as a couple of $M_J$
(e.g., Mall\'en-Ornelas et al.\ 2002).  Brown et al.\ (2001) give a
list of interesting follow-up studies possible for transiting planets
around bright stars.  For fainter stars large planetary moons could be
detected from transit timing.  More importantly, transit searches have
the possibility of probing new regions of parameter space compared to
current radial velocity planet searches.  For example, fainter stars
can be monitored in a photometric planet transit search, allowing
planets to be found in more distant environments and orbiting
intrinsically smaller stars.  Transit searches are unbiased with
respect to unusual spectral characteristics, which may lead to
unexpected discoveries and help constrain planet formation models.

One confirmed transiting extrasolar planet, HD~209458~b, is currently
known (Charbonneau et al.\ 2000; Henry et al.\ 2000).  HD~209458~b was
discovered by the radial velocity technique, and follow-up photometry
determined that it transits its parent star.  More than twenty groups
around the world are currently using photometry to search for
transiting extrasolar planets.  Many different environments are being
or have been searched, including the globular cluster 47 Tuc
(Gilliland et al.\ 2000), open clusters of different metallicities
(Howell et al.\ 1999; Mochejska et al.\ 2002; Quirrenbach et al.\
2001; Street et al.\ 2001, 2002; Burke et al.\ 2002), field stars with
magnitude $\lesssim 13$ (Borucki et al.\ 2001; Brown \& Charbonneau
2001), and $13 \lesssim I \lesssim 17.75$ stars in the direction of
the Galactic center (Udalski et al.\ 2002a,b).  There have also been
several conceptual papers about where and how to look for transits,
including early papers by Struve (1952), Rosenblatt (1971), and
Borucki \& Summers (1984), a thorough and prescient paper by Giampapa,
Craine, \& Hott (1995), and later papers after the discovery of CEGPs,
including cluster search strategies (Janes 1996) and searches towards
the Galactic Bulge with 10-m-class telescopes (Gaudi 2000).

Although no planets found from a transit search have yet been
confirmed with a mass detection from radial velocity follow-up, many
planet candidates now exist and are being followed-up for mass
confirmation.  Notably, the OGLE-III search (Udalski et al.\ 2002a,b)
has made public a list of over 50 stars with small transiting
companions, available to the astronomical community for follow-up.
The Vulcan project (Borucki et al.\ 2001; Jenkins, Caldwell \& Borucki
2002) and the EXPLORE project (this paper; Mall\'en-Ornelas et al.\
2002) have also produced planet candidates which are being followed-up
with radial velocity observations.  Although it is disappointing that
the transit search method has not yet resulted in any {\it confirmed}
planets, a steep learning curve as well as some lead-up time is
expected for this kind of enterprise.  It is not clear how much of the
current low planet yield results from the nature of the problem (e.g.,
due to an intrinsically low frequency of close-in planets in the types
of stars surveyed so far), and how much from technical challenges
faced by different transit surveys.  One of the limiting factors so
far is that many transit surveys have not monitored enough stars with
sufficiently high photometric precision.  This paper has the goal of
outlining the necessary steps for a successful transit search with a
CCD mosaic camera on a 2--4m telescope, and in addition describes the
main false positives that can be ruled out with high-precision, high
time sampling photometry and spectral typing of stars.

This paper presents a framework for the design of a search for
transiting planets around field stars and presents early results from
the EXPLORE (EXtrasolar PLanet Occultation REsearch) project.  The
EXPLORE project is a set of transit searches using wide-field mosaic
cameras on 4-meter-class telescopes with follow-up radial velocity
measurements on 8-meter-class telescopes.  We start in \S2 by
describing a very useful property of a high quality light curve with
two or more flat-bottomed transits: there is a unique solution for
planet and stellar parameters as long as certain important conditions
are met.  This unique solution can be used to obtain a clean set of
planet transit candidates and hence is one of the main motivations for
our experimental design.  When designing a transit survey it is
important to consider the frequency of transiting planets and the
transiting planet detection probability.  Section 3 presents a
description of the factors that affect the number of planets detected
and an estimate for the number of planets expected in a transit
survey.  The issues considered in \S3 are of a general nature and can
be applied to the design of any planet transit search.  When designing
a specific planet transit search, a number of interrelated choices
must be made in the observational design; in \S4, we continue the
discussion of survey design by describing the specific aspects of the
EXPLORE project strategy.  In \S5 we present early results from the
first search in our project, the EXPLORE I search, conducted in June
2001 at the CTIO 4-m telescope.  We summarize and conclude in \S6.
 
\section{The Unique Solution of a Two-transit Light Curve}
\label{sec-uniqueness}

One of the most attractive aspects of transit searches is that much
can be learned about a system with an orbiting companion from a good
light curve showing two or more eclipses.  Here we describe for the
first time that a light curve with two or more
flat-bottomed eclipses can in principle be used to derive a unique
solution of orbital parameters and companion radius, given certain
conditions.  The unique solution provides a powerful method to select
the best planet candidates to be followed up for mass determination.
Specifically, the stellar mass $M_*$, stellar radius $R_*$, companion
radius $R_p$, orbital distance $D$, and orbital inclination $i$ can be
uniquely derived from a light curve with two or more eclipses if the
following conditions are met:\\ $\bullet$ The light curve has
an extremely high photometric precision and high time sampling;\\
$\bullet$ The eclipses have flat bottoms (in a bandpass where limb
darkening is negligible), which implies that the companion is fully
superimposed on the central star's disk;\\ $\bullet$ There are no
secondary eclipses (i.e., the brightness of the companion is
negligible compared to the central star);\\ $\bullet$ The period can
be derived from the light curve (e.g., the two observed eclipses are
consecutive);\\ $\bullet$ The light comes from a single star, rather
than from two or more blended stars;\\ $\bullet$ The central star is
on the main sequence;\\ $\bullet$ The mass of the companion is
negligible compared to that of the central star ($M_p \ll M_*$);\\
$\bullet$ The orbit is circular (expected for CEGPs due to their short
tidal circularization timescales).\\

If the above conditions are met, the five parameters $M_*, R_*, R_p,
D,$ and $i$ can be uniquely derived from the five equations below.
For simplicity, the equations presented here assume that $R_* \ll D$.
The five equations are:

\noindent 
the transit depth
\begin{equation}
\label{eq:depth}
\Delta F = \left(\frac{{R_p}}{{R_*}}\right)^2;
\end{equation}
the relation between the inclination of the orbit and the shape of the
transit light curve, as parametrized by the ratio of the duration of
the transit's flat bottom $t_{flat}$ to the total transit duration $t_T$
\begin{equation}
\label{eq:shape}
\left(\frac{t_{flat}}{t_T}\right)^2 
= \frac{\left( 1 - \frac{R_p}{R_*}\right)^2 - \left(\frac{D}{R_*} \cos i\right)^2}
{\left( 1 + \frac{R_p}{R_*}\right)^2 - \left(\frac{D}{R_*} \cos i\right)^2};
\end{equation}
the total transit duration
\begin{equation}
\label{eq:duration}
t_T = \frac{P R_*}{\pi D}\sqrt{\left(1 + \frac{R_p}{R_*}\right)^2 - \left(\frac{D}{R_*} \cos i\right)^2};
\end{equation}
Kepler's Third Law
\begin{equation}
\label{eq:kepler}
P^2 = \frac{4\pi^2 D^3}{GM_*};
\end{equation}
and the mass-radius relation for (sun-like) main sequence stars
\begin{equation}
\label{eq:massradius}
M_* = f(R_*) \approx R_* \frac{M_{\odot}}{R_{\odot}}.
\end{equation}

The following observable quantities are measured from the light curve
and are used to solve the system of equations: the period $P$, the total
transit duration $t_T$, the flat eclipse bottom duration $t_{flat}$,
and the transit depth $\Delta F$.  Determining the
five parameters $M_*, R_*,R_p,D,$ and $i$ from the system
of equations and observables is a useful shortcut,
but in practice the final errors and parameters
are directly derived from a fit to the light
curve.  We note that five model parameters can be extracted from the
four observable quantities because of the assumed stellar mass-radius
relation (equation~(\ref{eq:massradius})), which provides a constraint
that does not depend on observations specific to a particular star.
Analyzing the error propagation through equations~(\ref{eq:depth}) to
(\ref{eq:massradius}) shows that the most critical inputs are
$t_{flat}$, $t_T$, and the mass-radius relation.  For system
parameters similar to those of HD 209458, errors of 10 minutes in
$t_{flat}$ or $t_T$, or a 20\% error in assumed radius at a given
mass, leads to errors of about 30\% in $M_*$ and $\cos i$, and about
15\% in $R_*$, $R_p$, and $D$.

\begin{figure}
\plotone{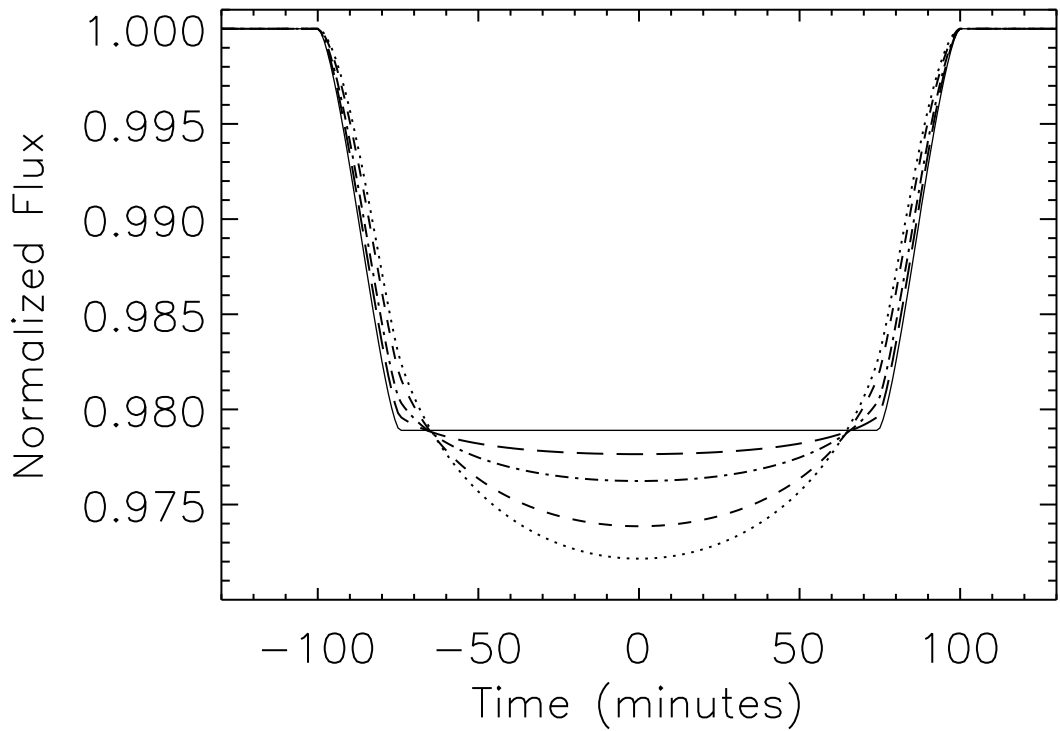}
\caption{Solar limb darkening dependence
of a central ($i=90^{\circ}$) planet transit light curve.  The planet
has $R=1.4R_J$ (approximately that of HD~209458~b) and the star has
$R=R_{\odot}$.  The solid curve shows a transit light curve with limb
darkening neglected.  The other curves, from top to bottom (at time =
0), show central transit light curves with solar limb darkening for
wavelengths (in $\mu$m): 3, 0.8, 0.55, 0.45.  Although the transit
depth changes at different wavelengths, the ingress and egress slope
do not change significantly.  The ingress and egress slope mainly
depend on the time it takes the planet to cross the stellar limb.}
\label{fig:limbdarkening}
\end{figure}

The presence of significant limb darkening will have an effect on the
transit depth and shape (see Figure~\ref{fig:limbdarkening}).  The
flat bottom of the transit will become progressively more rounded when
viewed at increasingly shorter wavelengths;  moreover, a central
transit will be deeper than $R_p^2/R_*^2$, since a larger fraction of
the stellar light is coming from a smaller area of the star.  The
shape of the transit can still be used to constrain the orbital
parameters if an appropriate limb darkening model is adopted.
However, given the extra parameters and the uncertainty in the adopted
limb darkening model, it is preferable to use light curves taken at
long wavelengths so that limb darkening is minimized
(Figure~\ref{fig:limbdarkening}).

If any of the assumptions listed at the beginning of this section are
incorrect for a given light curve, then the derived parameters ($M_*,
R_*, R_p, D,$ and $i$) will also be incorrect.  If $M_*$ and $R_*$ can
be obtained from spectral classification of the star, then the above
system of five equations and unknowns will be overconstrained and can
be used to check the correctness of the assumptions about the system.
For example, when two transits are separated by an odd number of
nights they might not be consecutive transits, since a transit may
have occurred during the day at time $P/2$.  In general, gaps in time
coverage can lead to missed eclipses and the resulting period
aliasing.  If period aliasing is suspected, the actual period can be
determined from the above system of equations plus a spectral type.
Even when only one high-quality transit is detected, the three
parameters $P$, $R_p$, and $i$ can still be constrained from
equations~(\ref{eq:depth}) through (\ref{eq:massradius}) provided the
spectral type---and hence $M_*$ and $R_*$---is known.  Another crucial
example of the usefulness of obtaining a spectral type is the case of
light curve from an unresolved triple system in which two of the stars
form an eclipsing binary system.  In this case, solving the above
equations under the assumption that the light is coming from only one
star will generally give a solution for $R_*$ and $M_*$ that is
inconsistent with the spectral type.  Thus, by complementing a
two-transit light curve with spectra or even broadband colors, much
can be learned about an eclipsing system, provided the photometry has
very high precision and high time sampling.  Further details, error
simulations, and applications of the unique solution to a transit
light curve are presented in Seager \& Mall\'en-Ornelas (2002).

\section{General Considerations for the Design of a Transit Survey}

In this section we discuss the factors that should 
motivate the basic design of any transit survey.  
The two most important broad considerations for 
designing a successful survey are:
(1) finding planets, and (2) providing useful statistics of planet
frequency and characteristics.
In broad terms,
the number of planets found by a transit search will be determined
by:\\ 
$\bullet$ frequency of
close-in planets around stars in the survey;\\ 
$\bullet$ probability of
having a geometric alignment that shows transits;\\ 
$\bullet$ number of stars surveyed;\\ 
$\bullet$ photometric precision;\\ 
$\bullet$ window function of the
observations.\\ 
The last three elements can be controlled by the survey strategy.  The
following subsections discuss the five factors listed above and their
significance for survey strategy.

\subsection{Planet Frequency and Detection Probability}
\label{sec:frequency}
An estimate of the fraction of field stars that have transiting
short-period planets is useful for designing a transit search.  The
frequency of transiting planets for a given ensemble of main-sequence
stars of similar metallicity, age, and environment can be
approximately written as

\begin{equation}
F_p  = \int \int \int P_p(R_*,R_p,D) P_g(R_*,D)
\hspace{0.1cm}dR_* dR_p dD.
\end{equation}
\noindent $P_p$ is the probability distribution that a star of
radius $R_*$ has a planet of radius $R_p$ with an orbital distance
$D$, and is precisely what a good survey should aim to measure.  $P_p$
is also likely dependent on stellar metallicity, age, and environment.
$P_p$ is currently not known because only a small number of
CEGPs have been found to date by radial-velocity (RV) planet searches.
$P_g$ is the geometric probability that a planet will occult its
parent star as seen from Earth; $P_g \sim R_*/D$ for an ensemble of
randomly oriented systems with circular orbits and $R_p \ll R_* \ll
D$.  A simple estimate of the frequency $F_p$ of transiting close-in
giant planets ($P\lesssim$4.5\, days) can be obtained by assuming all
isolated stars have the same frequency of CEGPs as isolated sun-like
stars, and adopting $P_p \sim 0.007$ (Butler et al.\ 2001) and the
corresponding $P_g\sim0.1$.  Assuming that we can detect planets only
around isolated stars, and adopting a binary fraction of 1/2, we get
$F_p=0.00035$.  In other words, we expect 1/3000 stars to have a
transiting close-in giant planet, with large uncertainty.  The
uncertainty comes from two sources. First, $P_p$ is measured only for
nearby isolated sun-like stars with planets of $M_p \simeq M_J$;
moreover this $P_p$ estimate comes from surveys that suffer from
limited statistics and selection effects which are difficult to
characterize. Second, it may be possible to detect transiting planets
around a star in a binary system, depending on the brightness ratio of
the stars and on the photometric precision.

In practice, the fraction of stars with planets actually {\it
discovered} by a transit search will likely be much less than $F_p
\sim 1/3000$.  The number of detected planets $N_p$ will depend
crucially on the window function of the observations $W$, the
photometric precision $\delta m$, the time sampling of the
observations $\delta t$, and the number of stars monitored $N(R_*)$.
The number of detected planets can be schematically written as
\begin{equation}
\label{eq:frequency}
N_p(W, \delta m) = \int \int \int N(R_*) P_p(R_*,R_p,D) P_g(R_*,D)
P_{det}(R_*,R_p,D,\delta m,\delta t) P_{vis}(W,D,R_*)
\hspace{0.1cm}dR_* dR_p dD.
\end{equation}
\noindent Here $P_{det}$ is the probability of detecting a transit of
depth $(R_p/R_*)^2$ given a photometric precision $\delta m$ and time
interval between photometric data points $\delta t$, and assuming the
transit occurs during the observations.  For a given photometric
precision, the significance of the detection will increase as the
square root of the number of photometric data points during transit;
this number depends on the time sampling of the observations $\delta
t$ and the transit duration (which is dependent on $D$, $M_*$, $R_*$,
$R_p$ and orbital inclination $i$, and will generally be close to 2--3
hours for close-in planets orbiting sun-like stars).  For a given
photometric precision, transits for a planet of a given size $R_p$
will be more easily detected around stars of smaller radius $R_*$.
Conversely, higher photometric precision will enable the detection of
smaller planets, or planets around larger stars (see
Figure~\ref{fig:paramspace}).  Note that a Jupiter-sized planet
transiting a sun-sized or smaller star will have transits of depth
$(R_p/R_*)^2 \gtrsim 1\%$, and will thus be easily detected in
well-sampled 1\% photometric precision light curves; i.e., for 1\%
photometric precision light curves, $P_{det}=1$ for Jupiter-sized
planets transiting sun-like or smaller stars.

\begin{figure}
\plotone{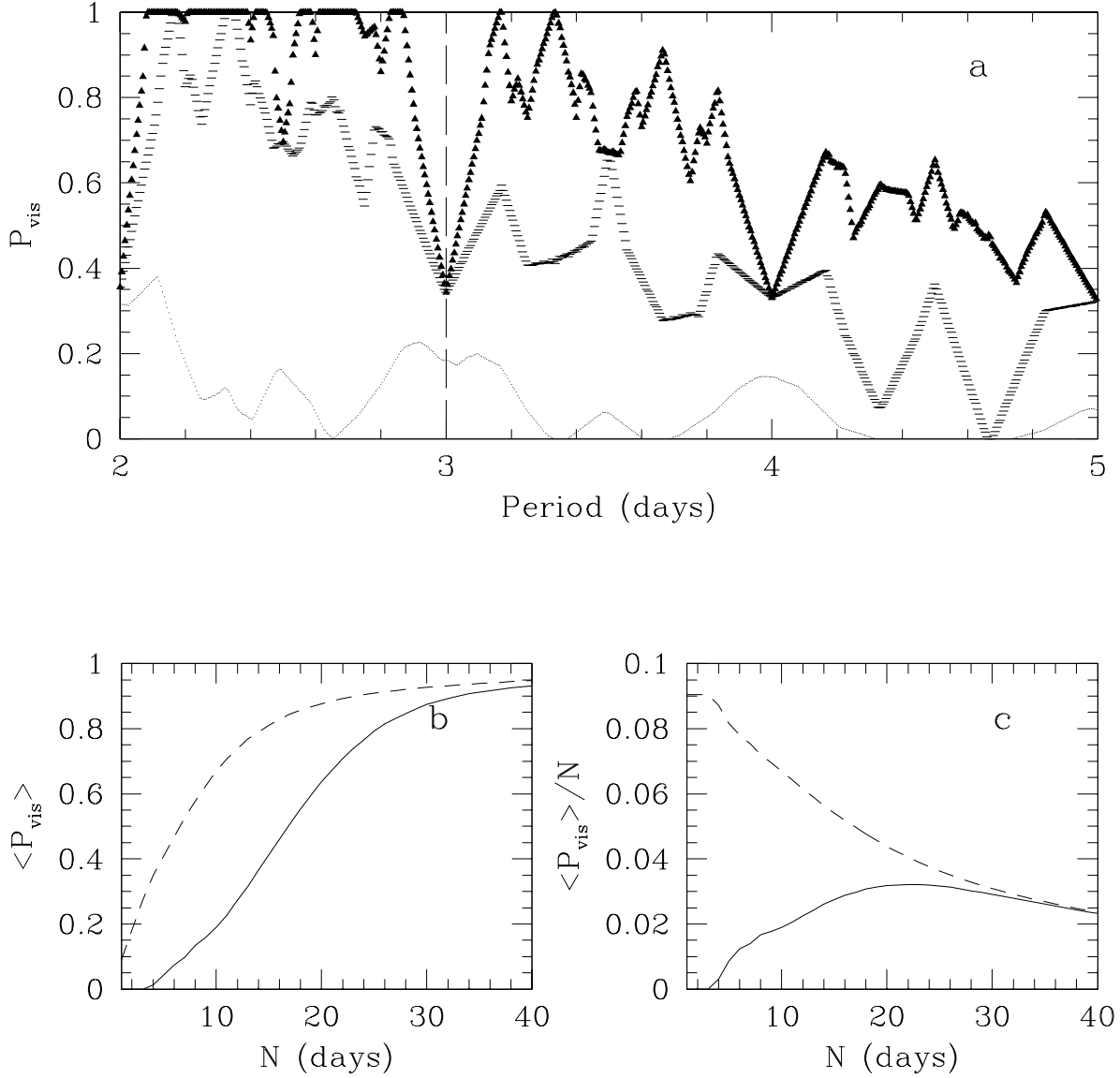}
\caption{Panel a: The probability $P_{vis}$ of detecting transiting 
planets with different orbital periods.  $P_{vis}$ is calculated with
the requirement that two transits must be observed.  Consecutive
nights of 10.8 hours each night are assumed.  At each period all
phases are considered; the difficulty of detecting some phases is
expressed by the dips in the curves.  For example, at integer periods
some transits will always occur during the day and will not be
detectable.  The different symbols are for transit searches of
different total number of nights: 21 nights (triangles), 14 nights
(bars), and the actual time coverage of the EXPLORE I search (dotted
line).  The vertical long-dashed line indicates lower period limit of
known CEGPs.  Panel b: The mean $<P_{vis}>$ as a function of number of
consecutive nights in an observing run.  The solid line is for the
requirement to detect two transits and the dashed line for one
transit.  Panel c: The efficiency of the $<P_{vis}>$ per night.  For a
two-transit requirement (solid line) the most efficient observing
length (for 10.8 hours each night) is around 21 nights but the
efficiency is similar for runs of 16 through 30 nights.  For a single
transit requirement the efficency curve (dashed line) decreases
monotonically because each additional night does not add new transit
detections, but the total $P_{vis}$ for one night is nonetheless tiny
(see panel b).}
\label{fig:visi}
\end{figure}

$P_{vis}$ is the probability that at least two full transits will
occur during an imaging campaign; observing at least two transits is
required in order to measure the orbital period and confirm the
transit.  $P_{vis}$ is a function of orbital period $P$, transit
length $t_T$, duration of observations each night, and number of
observing nights (described by the window function $W$).  Note that
$P$ and $t_T$ are in turn functions of $D, R_*$, and $M_*$ (or simply
$D$ and $R_*$ for main sequence stars).  Since we ultimately seek to
measure $P_{p}$, a good characterization of $P_{vis}$ is essential to
determine the frequency of planets around different types of stars.
Figure~\ref{fig:visi}a shows $P_{vis}$ for four different cases, all
with the requirement that two full transits are observed.  Shown are
periods of 2--5 days, although it should be noted that there are no
known extrasolar planets with periods below 3 days\footnote{Extrasolar
Planets Encyclopaedia, http://www.obspm.fr/encycl/catalog.html}.  The
case of $P_{vis}$ for 21 consecutive nights (shown by triangles) can
be used to illustrate the behavior of $P_{vis}$.  For planets of 2--3
day periods, most orbital phases would result in at least two transits
occurring during night-time over the 21 days of observations, and
the corresponding $P_{vis}$ is therefore unity.  Planets with longer
periods will have a smaller number of transits occur during the 21
day span of the run.  Consequently, as the period increases two transits
will be visible at night for a smaller fraction of orbital phases,
resulting in generally lower $P_{vis}$ for longer periods.  The
downward spikes in $P_{vis}$ at integer day periods (clearly visible
for the 21-day case) illustrate the limitations of a night-time
transit search done from a single observatory, since a percentage of
the transits with integer day periods will always occur during
daylight.  The effects of changing the length of the observing run are
illustrated by the other curves shown in Figure~\ref{fig:visi}a: the
bars correspond to 14 consecutive nights, and the dotted line
corresponds to the actual time coverage of the EXPLORE I transit
search at CTIO in June 2001 (11 nights, but only the equivalent of 6
nights had good weather).  Note that the $P_{vis}$ simulations
consider 10.8 hours of continuous observing each night.  However,
observing for 10.8 hours each night of the run may not be possible for
all combinations of field declination and observatory latitude, due to
the sliding of sidereal time throughout the run.  For an alternative
definition and discussion of $P_{vis}$ see Borucki et al.\ (2001)
and Giampapa et al.\ (1995).

When planning a transit search, it is important to consider the mean
planet detection efficiency $<P_{vis}>$ per observing run, and the
mean planet detection efficiency per night $\frac{<P_{vis}>}{N}$ as a
function of observing run length, as shown in Figure~\ref{fig:visi}b
and Figure~\ref{fig:visi}c.  The solid line in each graph corresponds
to the requirement that two transits are visible, while the dashed
line corresponds to one visible transit requirement and is shown as a
reference; in both cases $<P_{vis}>$ is calculated for planets with
3--4.5 day periods.  As one would expect, the mean planet detection
efficiency $<P_{vis}>$ increases monotonically as a function of
observing-run length.  This efficiency increase is very steep for runs
of $\lesssim$25 days.  The effective planet detection efficiency per
night, $\frac{<P_{vis}>}{N}$, has a broad peak around 21 days.  Thus,
for a site with perfect weather, it would generally be most efficient
to distribute observing runs in blocks of 3 weeks.  Note that
$\frac{<P_{vis}>}{N}$ decreases very sharply for runs lasting less
than one week.  Also note that for a single transit detection,
$\frac{<P_{vis,1}>}{N}$ is highest for the shortest observing runs,
and decreases monotonically for longer runs, since the extra nights
will result in repeat transit observations, which will not increase
$<P_{vis,1}>$ any further.

Based on equation~(\ref{eq:frequency}), our estimates of $P_{vis}$,
and our estimate of $F_p=0.00035$ for planets of 3--4.5 day periods,
we can calculate the expected number of transit detections in a
$\sim$2 week observing campaign (a reasonable limit for a shared
4-m-class telescope).  We estimate that during a perfect 13-night run,
$<P_{vis}>=0.3$, and thus one transiting close-in giant planet will be
discovered for every $\sim10,000$ stars with $\lesssim 1\%$ precision
light curves.  For a perfect 17-night run, $<P_{vis}>$ increases to
0.5, and one transiting planet detection should be expected for every
$\sim 6,000$ stars observed with $\lesssim 1\%$ photometric precision.

\subsection{Maximizing $P_{vis}$}
\label{sec-Pvis}

Maximizing $P_{vis}$ is a main consideration for maximizing observing
efficiency.  A useful transit search will be one that produces a clean
set of candidates with minimal contamination from false positive
planet detections.  In this section we consider possible strategies
for allocating a given limited number of observing hours in the
context of trying to produce the cleanest set of planet candidates.
Strategies can range from carrying out a few observations per night
over a large number of nights, to carrying out all observations in a
single observing run of consecutive nights.  We will focus on a
scenario in which the telescope is not dedicated to transit searches
(e.g., a shared national or international facility).  We argue that
when observing time is limited to a few weeks in a given season, the
cleanest set of transit candidates will be obtained by conducting
observations in one contiguous block.

A transiting close-in planet with a period of $\sim$3--4 days will
typically have very few transits occurring at night during a 2--3 week
observing campaign.  A well-sampled eclipse light curve and reliable
detection is most easily achieved with high cadence observations in
which an eclipse can be detected in a single night (i.e., without
folding the light curve).  In principle it is also possible to detect
the 1\% dips in the star's flux caused by a transiting planet as a
periodic signal in sparse observations done over many nights or
even weeks, as long as extremely high photometric precision is
achieved from night to night.  Unless the observations have a very
long baseline and many hundreds of data points are obtained for each
star, the resulting phased light curve will most likely not have
enough in-transit data points to enable a measurement of the shape of
the eclipse.  Knowing the shape of the eclipse is critical for ruling
out common contaminants with periodic 1\% dips, such as grazing
binaries (see
\S\ref{sec:contamination}; Seager \& Mall\'en-Ornelas 2002).  Such
contaminants can introduce many false positive detections to the list
of planet candidates, making follow-up very inefficient and
statistical analysis very difficult. Therefore when only a few weeks
of telescope time are available, it is best to obtain high cadence
observations in which well-sampled transits are detectable in a single
night.

\begin{figure}
\plotone{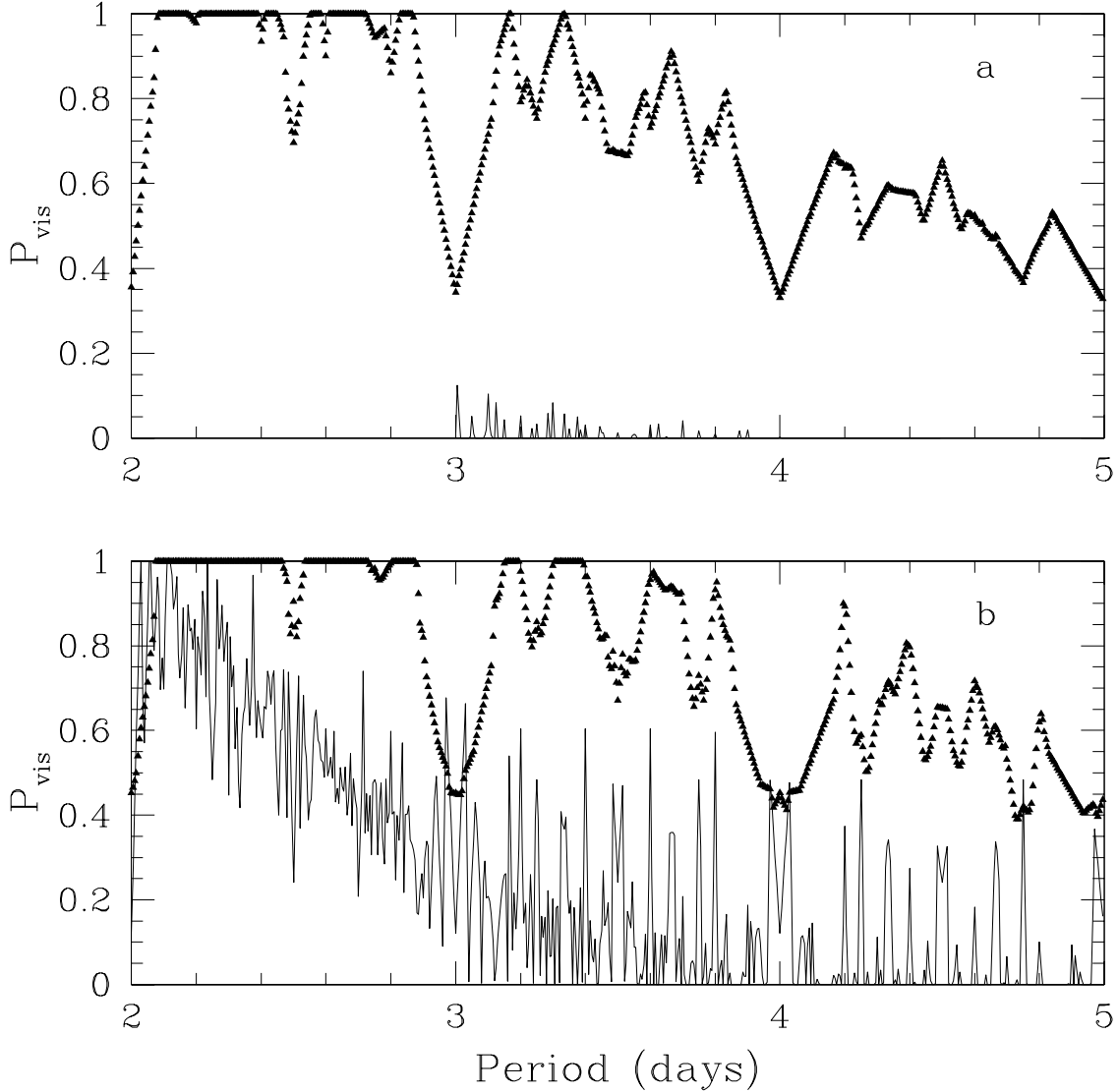}
\caption{A comparison of $P_{vis}$ for two different observing
strategies: full night or partial night observations.  The triangles
show the fiducial model: $P_{vis}$ for 21 consecutive nights with 10.8
hours each night (the same curve as in Figure~\ref{fig:visi}a).  Panel
a shows $P_{vis}$ from the fiducial model compared to $P_{vis}$ from a
transit search with the same two-transit requirement and the same
total observing time spread out over 76 nights with 3 hours per night.
Almost no full transits are detected twice in the partial night
piece-wise strategy.  Panel b shows the same comparision as in panel a,
but with the requirement relaxed to detect 4 half transits.  Under the
relaxed requirement, both search strategies will detect a larger
percentage of the existing transits.  However, the piece-wise search
will still find far fewer transits than the 21-night search.  As
discussed in the text, the piece-wise transit search strategy has
additional serious practical problems such as false partial transits
caused by systematic errors in the photometry.}
\label{fig:visibad}
\end{figure}

If a transit is to be detected within a single night, it is best to
detect a full transit rather than a partial transit, since there are
common systematic errors in the photometry which can mimic the
beginning or end of a transit.  Full transits are best for determining
transit length and shape, which are necessary to constrain system
parameters and find good planet candidates for follow-up
(\S\ref{sec:contamination}).  Full transits are only visible when the
middle of the transit is within the middle ${\rm L}-t_T$ hours of the
observations, where L is the number of hours of continuous
observations and $t_T$ is the total transit length in hours.  This
implies that observations should be taken for as many continuous hours
as possible.  For example, consider a series of observations lasting 4
hours at a time.  Such a strategy would be extremely inefficient for
finding transits, since a 3-hour transit (the typical length for a
close-in planet around a sun-like star) could only be detected if it
was centered during the middle hour, i.e., just 25\% of the observing
time.  In contrast, the transit detection efficiency for a series of
11-hour observations is much higher, since a 3-hour transit would be
detected if centered in the middle 8 hours of observations, or 73\% of
the observing time.  Figure~\ref{fig:visibad} illustrates the
importance of scheduling full nights of observations by comparing
$P_{vis}$ for 227 hours of observations scheduled in either 21 full
nights of 10.8 hours, or 76 nightly 3-hour segments.

It is best to schedule all nights in a contiguous block so that the
field of choice is visible for as long as possible each night
throughout the run; if the allocated nights are spread out over many
weeks, changes in sidereal time will cause the field to be visible for
only part of the night, thus decreasing $P_{vis}$.  In an ideal case
observations would be done from several observatories spread out in
longitude, or from a location where it is possible to obtain
continuous coverage of the field.
Period aliasing will be a severe problem if a given number
of nights are split into two observing runs separated by a year
instead of in one long block, since the period will be effectively
unknown for any systems that have only one transit detected in each of
the observing runs.  Without a period determination, the
characteristics of the eclipsing system will be unknown, and the
interpretation of a radial-velocity follow-up (with only a few data
points) will be severely limited.

We have argued in favor of allocating observing time in one contiguous
block, which has the advantage that a higher photometric precision can
be achieved in a single night rather than across several different
nights.  This strategy is well suited to the constraints of using
large telescopes in shared national or international facilities, where
a realistic time allocation in one season is limited to 2--3 weeks.
Note that once the best planet candidates have been identified, it is
very useful to conduct additional observations in short sets spread
out over several weeks or months in order to obtain an accurate period
measurement and confirm the transits.

An alternative strategy to the one we advocate above would be to take
less-frequent observations of the field over many weeks or months, and
fold the light curve with a variety of periods in order to search for
transits.  This strategy is most suitable for private telescopes which
can be dedicated to a photometric campaign over many weeks.  By
observing the field for many seasons, the total number of photometric
data points taken in this way would eventually be the same as in the
strategy with contiguous high time sampling observations, resulting in
the same effective time sampling.  This strategy can gain statistical
certainty by summing phased data from different transits, but it
requires extremely high photometric precision from night to night.
The advantage of having sparse observations over a long baseline,
however, is that the period can be determined to very high accuracy.
For a good example of this strategy see the OGLE-III planet transit
search (Udalski et al.\ 2002a).

\section{A Field Transit Survey Design: the EXPLORE Project}

The general framework presented in the previous section can be used in
the design of any transit survey.  In order to design a specific
transit search, a large number of interrelated issues must be
considered.  In this section we discuss many relevant factors which
affect transit survey design, and describe the specific choices made
by the EXPLORE project.  The EXPLORE project is a series of searches
for transiting planets around Galactic plane stars.  The main
considerations for the EXPLORE program design are to maximize
$N(R_*)$, $P_{det}$, and $P_{vis}$ (see
equation~(\ref{eq:frequency})), and to minimize false-positive
detections to obtain a high yield of actual planets among the transit
candidates.  We maximize $N(R_*)$ by using 4-m-class telescopes with
large-format CCD mosaic cameras to look in the Galactic plane.  We
maximize $P_{det}$ and minimize false positive detections by carrying
out high-precision photometry with high time sampling on a single
field with mostly main sequence stars.  We maximize $P_{vis}$ by
monitoring the selected field for as many consecutive nights as
possible, for as long as possible each night.  As part of the program
design, we reduce the data and find planet candidates within a few
weeks of the observations; this allows us to follow up planet
candidates with radial velocity measurements in the same season,
before orbital phase information is lost.  The following subsections
explain the details of the EXPLORE project experimental design.

\subsection {Instrument Selection} From the estimates of transiting
planet frequency and $P_{vis}$ presented in \S\ref{sec:frequency}, it
is clear that many thousands of stars must be monitored in order to
detect a single planet.  The most efficient approach is to use a
wide-field instrument to observe an area in the sky with a high
density of stars.  Note that the total number of pixels in the
detector is crucial for determining how many stars can be
monitored with high photometric precision ($< 1\%$).  The relationship
between the field size and the telescope aperture and instrument
efficiency is also very important, since a high time sampling
is essential in order to pick the best planet candidates without major
contamination from grazing binaries and blended stars (see
\S\ref{sec:contamination}).  The EXPLORE project currently has 
searches using the MOSAIC II camera at the CTIO 4-m telescope
(8K$\times$8K pixels, $36'\times36'$ field of view), and the CFH12K
camera at the 3.6m CFHT (8K$\times$12K pixels, $28'\times42'$ field of
view).  Both searches have a high time sampling with photometric
measurements every $\lesssim$3 minutes.

It is useful to outline some benefits of using a large telescope to
conduct a narrow-angle transit search on relatively faint stars.
First, deep transit searches contain a proportionately larger number
of low-mass, low-luminosity stars than shallow searches; in
particular, deep transit searches probe intrinsically fainter stars
than most ongoing RV searches.  For a search conducted with a
large-format mosaic CCD camera on a 2--4 meter telescope, a very large
number of stars can be observed at once with good photometric
precision and time sampling. The large number of stars directly
increases the chances of finding a transiting planet.  Observing many
stars at once also helps improve the the precision of the relative
photometry.  The PSF is usually well sampled in the case of large
telescopes, which results in higher photometric precision than is
possible with the typically under-sampled PSF of small telescopes with
large fields of view.  Furthermore, there is less concern about
possible systematic errors introduced by differential extinction
across the relatively small field of a deep survey than in the
large-angle field of a small telescope.  Finally, going to fainter
apparent magnitudes reduces the proportion of intrinsically bright and
distant giant stars in the observed sample, leaving a larger fraction
of main-sequence stars useful for finding planets.  Conversely, there
are two main disadvantages associated with using a large telescope:
(1) is more difficult to get large amounts of telescope time to
improve time coverage, and (2) since the stars are fainter, an 8--10m
class telescope is required for the RV follow-up, and the range of
feasible follow-up studies is much more limited than for bright stars.

\subsection {Field and Filter Selection}

Lower main sequence stars are the only stars in which transits by
Jupiter-sized planets are easily detected, since for larger stars the
$(R_p/R_*)^2$ dip caused by a transiting planet will be much smaller
than 1\%.  In this section we discuss our choices of field and
filter, both aimed at obtaining the largest number of lower main
sequence stars observable with better-than-1\% photometric precision.

In order to get a large number of lower main sequence stars we look at
the Galactic plane.  With large-format CCD mosaic cameras on 4-m-class
telescopes, it is possible to find Galactic plane fields with
100,000---500,000 stars detected in 1--2 minutes of integration.  The
initial consideration for picking a field is the time of the year, since
the Galactic plane must be visible from the observatory in question.
For a given site, the best combination of long nights and good weather
is important for choosing the best month for the observations.  It is
also important to have the Galactic plane at a declination which will
make it visible for as many hours as possible in a given night.  After
observing time is allocated, we preselect an area in the Galactic
plane by considering the following: (1) sidereal time matched to 
RA so that the field transits the meridian in the middle of the night
at the center of the observing run; (2) low dust content based on the
dust map by Schlegel, Finkbeiner \& Davis (1998) and the CO map by
Dame, Hartmann, \& Thaddeus (2001); (3) high stellar number counts
from the USNO2 catalogue (Monet et al.\ 1998) and the Digitized Sky
Survey 2\footnote{The Digitized Sky Survey was produced at the Space
Telescope Science Institute under U.S.  Government grant NAG W-2166.}.

\begin{figure}
\plotone{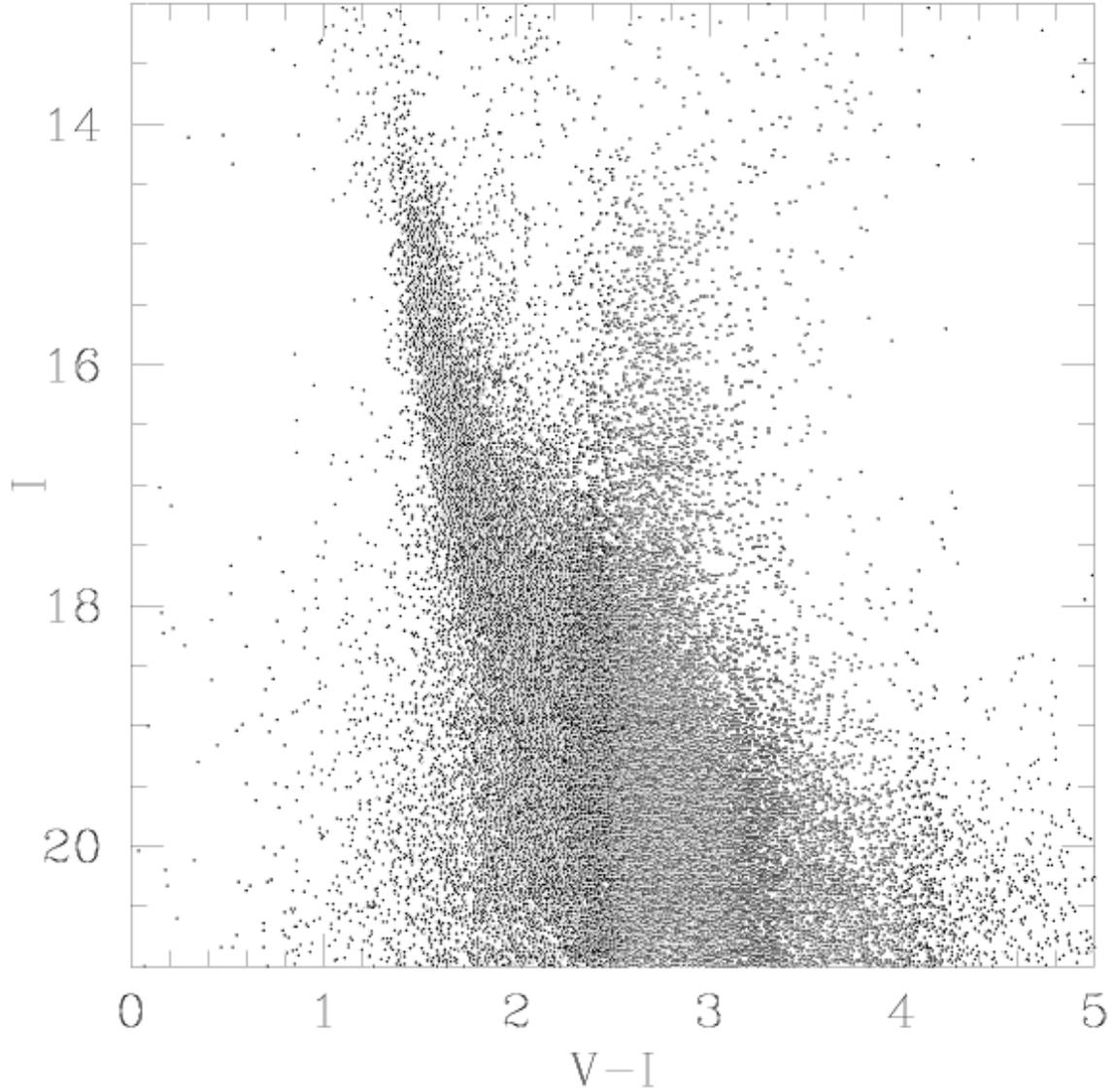}
\caption{Color-magnitude diagram
for chip 1 from the EXPLORE I field.  The giant branch and main
sequence are clearly visible, showing that the fraction of giant stars
is small.  Most stars fainter than magnitude $I=18$ are too faint for
$<1$\% photometry, but faint star data points in the light curves can
be co-added to get better precision.}
\label{fig:CMdiagram}
\end{figure}

\begin{figure}
\plotone{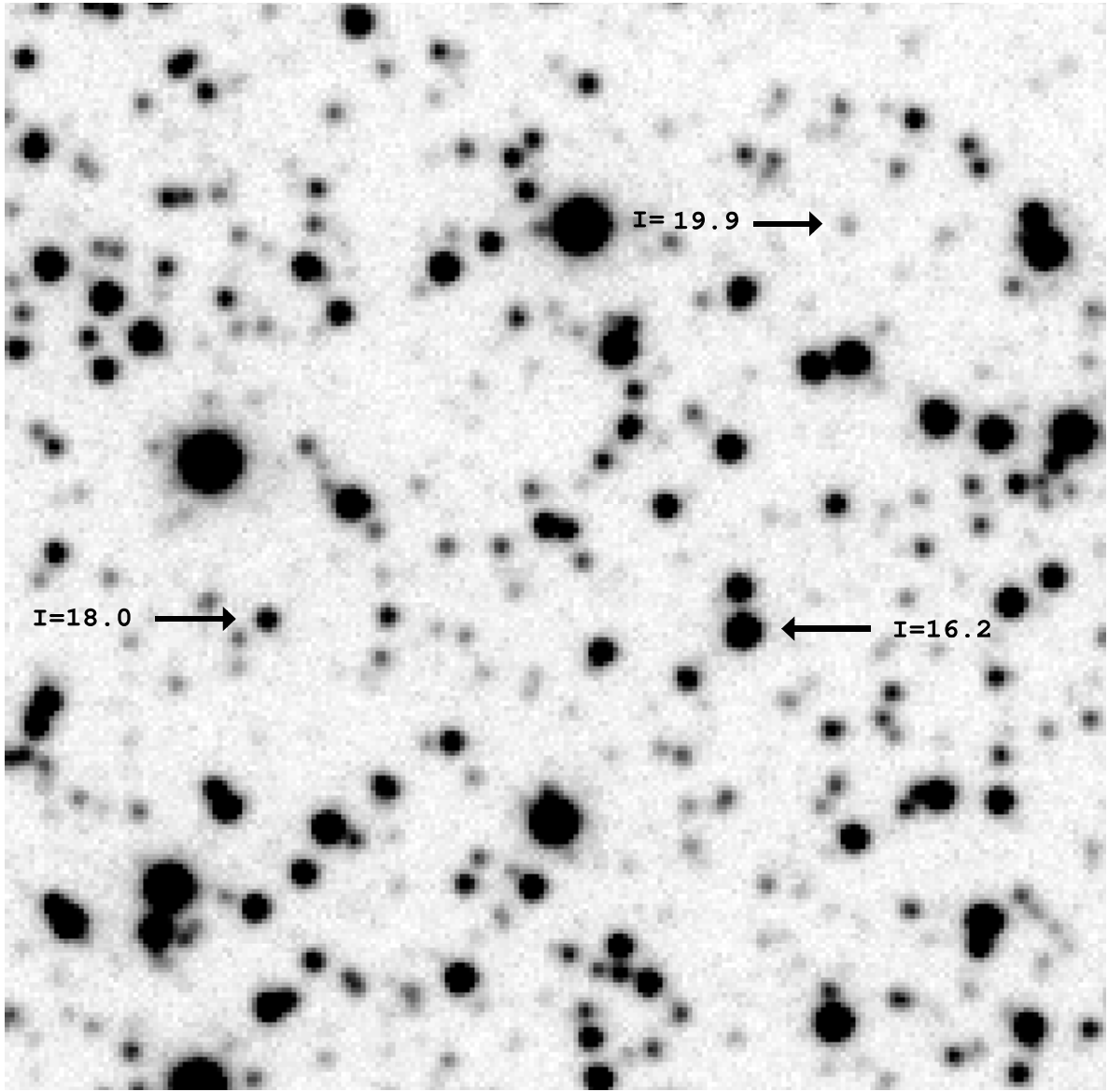}
\caption{A sample 60$''\times60''$ square of a
survey image.  The whole MOSAIC II field is $36'\times36'$, or $\sim$1400
times this area.  
Note that a large fraction of the stars are relatively
isolated, and a significant fraction of the image
area is free from stars, which still permits
a good determination of the sky level.
The arrows point to three stars
with different magnitudes.  The two at $I$ = 16.2 and 18.0 span the
range of most of our stars suitable for high-precision photometry.  The
star with $I$ = 19.9 is too faint for $<$ 1\% photometry in a single
frame.}
\label{fig:image}
\end{figure}

In order to choose the best field within the preselected region we
take $BVRI$ test images of several fields, using the same instrument
to be used for the actual search.  We compute number counts and
construct color-magnitude (CM) and color-color diagrams for each
field, and choose the field with the best combination of the following
factors: highest proportion of lower main-sequence stars, uniform and
low dust extinction, and smallest number of bright stars that saturate
large areas in the CCD.  As an example, Figure~\ref{fig:CMdiagram}
shows a color-magnitude diagram of one chip in the field of the
EXPLORE~I search at the CTIO 4-m telescope.  An additional
consideration could possibly be to choose an uncrowded stellar field
so that that the photometric precision is not adversely affected by
crowding.  In practice, however, most Galactic plane fields are not
significantly crowded for the 1--2 minute exposures and 4-m-class
telescopes used in the EXPLORE project.  Moreover, effective
photometry algorithms have been developed to handle crowded field
relative photometry (e.g., difference imaging by Alard \& Lupton 1998;
Wo\'zniak 2000; Udalski et al.\ 2002a).  Figure~\ref{fig:image} shows
an example of a small region (1/1400 of the total area) in the
EXPLORE~I field, which is located in the Galactic plane at $l=-27.8$,
$b=-2.7$.  Note that a large fraction of the stars are relatively
isolated, and a significant fraction of the image area is free from
stars, which permits a good determination of the sky level (crucial
for faint star photometry).

An important consideration is to minimize the number of giant stars,
since they are too large to be useful for planet transit detection
(since $(R_{J}/R_*)^2\ll 1\%$) and are a major source of contamination in
shallow (mag~$\lesssim$~13), wide-field transit surveys (W.  Borucki
2001, private communication; D.  Latham 2001, private communication).
In the case of the EXPLORE project, we minimize the proportion of
contaminating giant stars in our sample by observing the Galactic
plane with deep images (e.g., $15 \lesssim I \lesssim 20$).  A giant
star would have to be nearly outside the Galaxy in order to have the faint
apparent magnitude of most stars in our survey; for example, a K5
giant with an apparent magnitude of $I\sim17$ (assuming 1 magnitude of
extinction in $I$) would be at a distance of 54 kpc, where the Galactic
stellar density is extremely low.  Also, we select fields away from
the Bulge in order to avoid bright giant stars, which would often
saturate the CCD and increase crowding in the field.

Using a very red filter allows us to maximize the number of lower
main-sequence stars (i.e., relatively small stars), for which transits
are most easily detected for a given size planet.  Specifically, an
$I$-band filter increases the counts of stars of later type than the
sun by a factor of 2 to 6 over that of $R$-band (for a fixed magnitude
range).  Observing in the $I$ band minimizes the effects of absorption
by interstellar dust as compared to bluer bandpasses.  Finally, the
choice of the $I$ band produces light curves with the least
significant limb-darkening among standard $BVRI$ filters (see
Figure~\ref{fig:limbdarkening}); this is extremely useful for
selecting eclipses with clear flat bottoms and for deriving the best
transit parameters without significant dependence on uncertain
limb-darkening models (see \S\ref{sec-uniqueness}).

\subsection{Photometric Precision and Time Sampling}
\label{sec:photpipe}

High photometric precision and high time sampling are crucial in order
to identify the best set of transiting planet candidates with minimal
contamination, as was described in \S\ref{sec-uniqueness}.  In the
EXPLORE project, we achieve both high photometric precision and high
time sampling by monitoring a single field throughout the observing
run, with exposures taken every $\sim3$ minutes.  Care is taken so
that the field position does not shift in the CCD by making small
adjustments to the pointing throughout the night, and average net
shifts in the field position are kept to $<1''$.  This is done in
order to minimize photometric errors introduced by residual
differences in the CCD response across the chip that are not
completely taken away by the flat-field correction, and to simplify the photometry pipeline algorithm.

Observing a single field is the strategy which achieves the highest
time sampling.  A more complicated strategy in which several fields
are monitored at once by switching from field to field would
significantly decrease time sampling of each field.  Also, setting up
the position and guiding of each field many times throughout the night
would likely lead to a large waste of observing time.  Another
alternative strategy of switching fields only once or twice throughout
the night would mean that each field would only be observed for 3--5
hours at a time; this would significantly reduce the transit detection
efficiency $P_{vis}$, as discussed in
\S\ref{sec-Pvis}.

\begin{figure}
\plotone{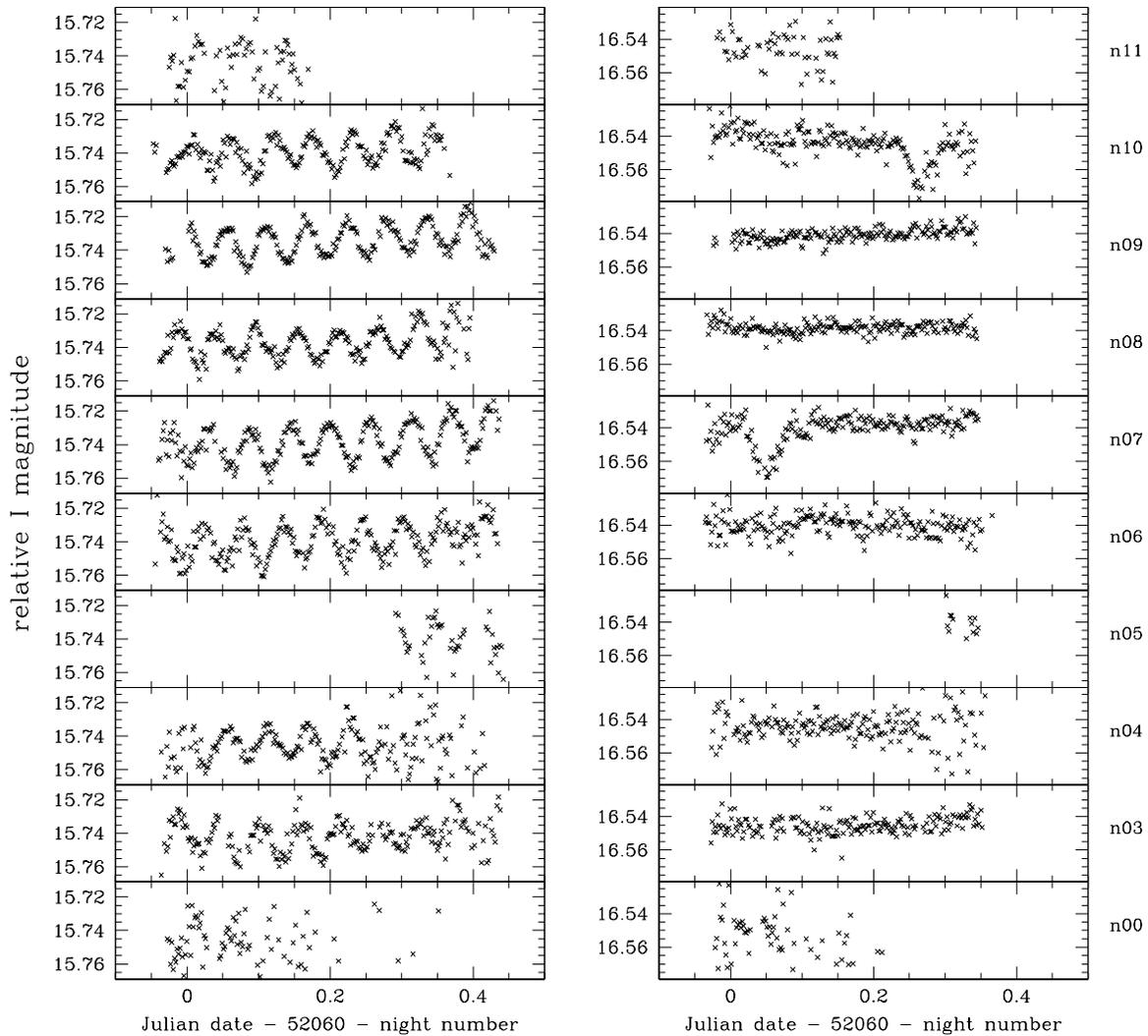}
\caption{Two examples of the high time sampling and
high photometric precision light curves from the EXPLORE I search.  The
rows correspond to different nights of data where the night numbers
are listed at the far right of the figure.  Note that the photometric
precision varies from night to night and on some nights is especially
poor due to clouds and bad seeing.  The dome was closed on a large
portion of nights 1, 5, and 11, and was closed for all of night 2.
There is a one night gap in the time assignment between nights 1 and
2.  Panel a: Many low-amplitude variable stars, such as this $\delta$
Scuti star, are found in our data set.  Panel b: A light curve of a
likely grazing eclipsing binary star from the EXPLORE I search.  The
round bottom and very sloped ingress and egress are indicative of a
grazing binary system.  The photometric precision and high time
sampling of our data are good enough to rule out grazing binary stars,
a common contaminant in other transit surveys (W.  Borucki 2001,
private communication). Notice the scale on the $y$ axis, which
clearly shows that our relative photometry reaches a precision of
considerably better than 1\%.}
\label{fig:grazing}
\end{figure}

We have developed a customized pipeline to perform high-precision
photometry of faint stars in dense fields with a well-sampled PSF.
Full details of our photometry pipeline will be described in Yee et
al.\ (2002), and we present a brief summary of the algorithm here.  A
key feature of our high-precision photometry algorithm is the use of
relatively small apertures (about a factor of two to three larger than
the seeing disk, i.e., a diameter of 2$''$ to 3$''$) for measuring the
flux.  This stems from the requirement to minimize the contribution of
sky noise for stars that are not significantly brighter than the sky
(as faint as $I\sim19$).  The crucial consideration in obtaining 
high-precision relative photometry when using such small apertures is the
exact placement of the center of the aperture relative to the centroid
of the stars.  We achieve this high-precision aperture placement by
using an iterative sinc-shift algorithm to resample {\it each} star so
that the central $3\times3$ pixels sample the PSF symmetrically (Yee
1988).  A photometric growth curve for each object is then derived
using integer pixel apertures on the resampled image of the star.  The
resampling is equivalent to placing all the stars within the
photometry aperture in an identical manner, allowing for relative
photometry to be carried out using much smaller apertures than is
customarily done.  The photometric measurements of each star are then
put on a relative system by comparing them to a set of reference stars
determined using an iterative algorithm to find the most stable stars
in a given region of the CCD.  Light curves are finally produced
based on the relative photometry.  Examples of the high-quality light
curves achieved using the first version of our pipeline are shown in
Figures \ref{fig:grazing} and \ref{fig:4343}, and further discussed in
\S\ref{sec-light curves}.

\subsection{Data Reduction Strategy}
\label{sec-strategy}
A key feature of the EXPLORE project is that the data reduction and
analysis is done on a short timescale.  In order to produce light
curves within 1--2 weeks of the end of an observing run we have
developed a pipeline which runs on a dedicated computer cluster.  Our
pipeline consists of custom-written programs to do image
pre-processing, aperture photometry, relative photometry, and to
generate light curves.  The only steps that currently require
significant human intervention are visual verification of the
automatic object finding performed using the program PPP (Yee 1991) to
create a star catalog, and finding the best parameters for the
relative photometry.  The latter step will eventually be done
automatically as well.  The main bottleneck that currently prevents us
from reducing data in real time (which is our eventual goal) is the
long time it takes to read the raw data tapes written at the telescope
into the computer cluster where the data are reduced.  A main
motivation for the fast data reduction is that follow-up radial
velocity observations of transiting planet candidates are best
interpreted if done in the same season when the phase of the orbit is
known.  For a two week or shorter observing run, the baseline for
determining the period is small so that typical errors in the period
will accumulate over a year and the phase will likely be lost.

\subsection {Follow-up Radial Velocity Measurements}
\label{sec-rv}

Late M dwarfs ($M \geq 80 M_J$), brown dwarfs ($13 M_J < M < 80 M_J$),
and gas giant planets ($M \leq 13 M_J$) are all of similar sizes due
to a competition between Coulomb force effects ($R \sim M^{1/3}$) and
electron degeneracy pressure effects ($R \sim M^{-1/3}$) (Hubbard,
Burrows, \& Lunine 2002).  Hence, transits alone are not enough to
determine that a transiting companion is actually a planet even if the
radius is constrained to be $\sim$ 0.1--0.15 $R_{\odot}$.  Radial
velocity (RV) measurements therefore are needed to determine the mass,
and thus the nature, of the orbiting companion.  RV measurements are
also useful to rule out grazing binaries and other possible
contaminants that mimic the transit signature, which can be common in
the case of noisy light curves (\S\ref{sec:contamination}).  The
transit search method with follow-up radial velocity confirmation is
very powerful because every planet found has a measured radius and an
absolute mass.  Obtaining a mass measurement for transiting planet
candidates is facilitated by the fact that the orbital period and
phase are known a priori, and therefore observations can be conducted
at pre-determined times such that the radial velocity differences are
maximized.  In the case of faint stars where planet masses lower than
1--2 $M_J$ cannot be detected due to limitations in the currently
achievable RV precision, it is still possible to determine a mass
upper limit of a few $M_J$ (e.g., Mall\'en-Ornelas et al.\ 2002).  An
actual mass measurement is generally required for confirmation of the
presence of a planet.  However, a mass upper limit could still be used
to make a case for the presence of a planet {\it as long as all
possible contaminants to the transit signature can be ruled out with
confidence} (see \S~\ref{sec:contamination}).

\begin{figure}
\plotfiddle{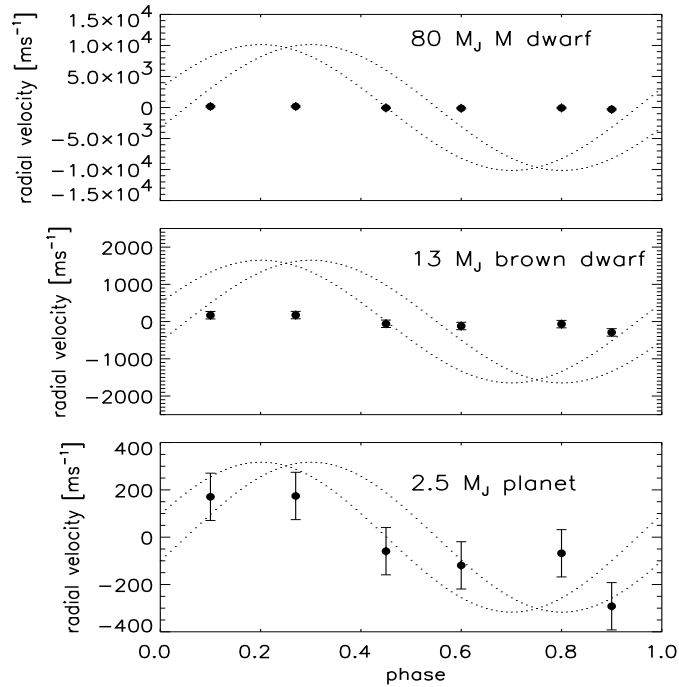}{4in}{0}{75}{65}{-220}{-170}
\caption{Model radial velocity
curves (dotted lines) for companions of different mass orbiting a
solar-mass star with $D=0.05$~AU (corresponding to $P=4.08$ days).
Overplotted are simulated radial velocity data points for a 2.5 $M_J$
companion, with 100 m/s rms noise and well-spaced in phase.  The
adopted error bar is 100 m/s, approximately what is attainable with
hour-long exposures on faint stars ($V \sim 18$) with 8m-class
telescopes (Mall\'en-Ornelas et al.\ 2002).  The two dotted lines in
each panel show the uncertainty in orbital phase corresponding to
accumulated 20-minute errors in the 4-day period over 4 months.  It is
evident from the top two panels that even with a radial velocity
precision of $\sim$ 500 m/s, transits due to a stellar companion or
brown dwarf companion can be easily ruled out.  The $y$ axes are
different for each panel; the dotted lines look similar in each panel
because the radial velocity amplitude scales linearly with companion
mass (equation~(\ref{eq:rv})).}
\label{fig:rv}
\end{figure}

The amplitude of the RV variations of a star in the presence of a less
massive companion in a circular, edge-on orbit is:
\begin{equation}
\label{eq:rv}
K = 2 \pi \left( \frac {G}{4 \pi^2 P} \right) \frac{m_2}{m_1^{2/3}}.
\end{equation}
Here $P$ is the period and $m_1$ and $m_2$ are the primary and
secondary masses, respectively.  Because transiting planet orbits are
seen almost completely edge-on ($i \sim 90^{\circ}$), the full RV
variation is along the line of sight.
A G2V star ($M=M_{\odot}$) with an 80 $M_J$ M dwarf or a 13 $M_J$
brown dwarf companion with an orbital distance $D=0.05$~AU
(corresponding to $P=4.08$ days) will show RV amplitudes of 10.1 km/s
and 1.6 km/s, respectively.  Thus both M dwarfs and brown dwarfs are
very easy to rule out with $<$ 500 m/s RV precision (even for stars
more massive than G2V and for stars with planets in slightly longer
period orbits).  A simulated example is shown in Figure~\ref{fig:rv}.
Note that a radial velocity precision of 500 m/s is easily attainable
with an echelle spectrograph on an 8-m-class telescope even for the
relatively faint stars ($I\lesssim18$) in the EXPLORE project.

\begin{figure}
\plotfiddle{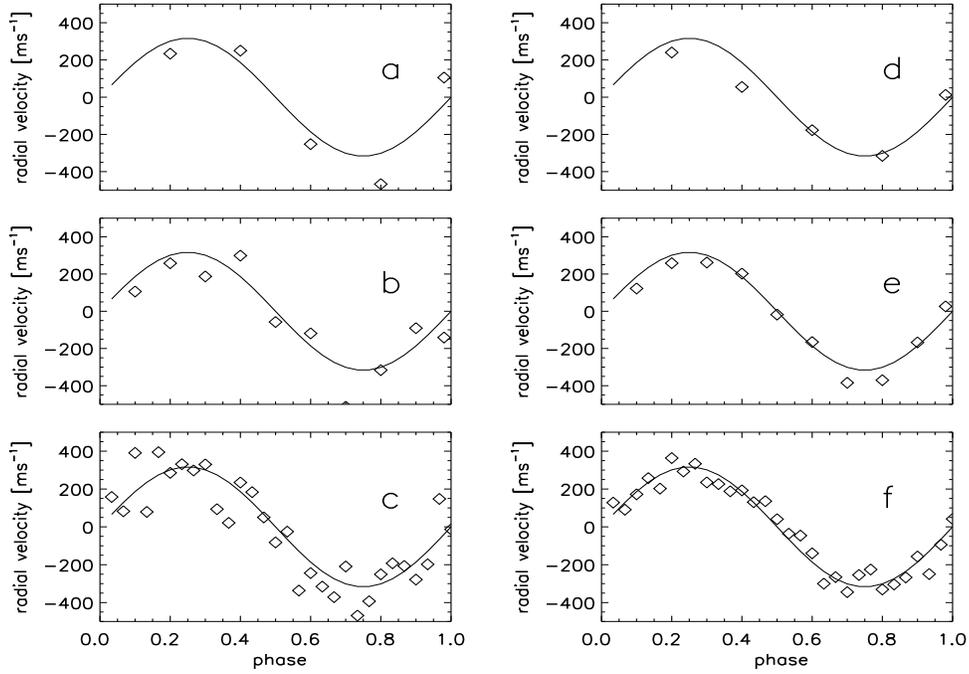}{4in}{0}{75}{65}{-220}{-170}
\caption{A simulated example showing that only a few radial
velocity measurements are needed to constrain the mass of a close-in
extrasolar giant planet if the phase of the eclipsing system is known.
The solid lines show a model radial velocity curve for a 2.5 $M_J$
planet orbiting a solar-mass star at 0.05 AU (which corresponds to a
4.08 day period).  Panels a--c show theoretical radial velocity points
with added Gaussian noise of $\sigma = $100 m/s.  Shown in each panel
are 5, 10, and 30 radial velocity points, respectively.  Panels d--f
show the same as panels a--c but with a noise of 50 m/s.}
\label{fig:rvnights}
\end{figure}

We note that planet searches that use RV measurements to find planets
reach a precision of a few m/s (e.g., Butler et al.\ 1996; Pepe et
al.\ 2000).  This level of precision is important when one is trying
to detect possible periodic changes in the RV and measure orbital
parameters, but is not necessary when trying to distinguish variations
of widely different amplitudes for a system with a known period and
phase.  Radial velocity follow-up confirmation of transit candidates
can be extremely efficient.  As shown in Figure~\ref{fig:rvnights},
only a handful of RV points at a judiciously chosen time are needed to
constrain the companion's mass, as long as the period and phase are
known.  Knowing the transiting companion's orbital phase is very
important when interpreting the RV measurements.  A small error in the
period measurement from a two-transit discovery light curve will
rapidly accumulate with each orbit to give a phase error that
increases linearly with time.  For instance, in one year a planet with
a $3\pm 0.007$ day period (i.e., a 10-minute uncertainty) will have an
accumulated error of 0.85 days, or 0.3 in phase.  Thus, for a
$\sim$two-week observing run (with only a short baseline for period
determination) it is best to do follow-up observations in the same
season the discovery light curve is taken, since otherwise a second
imaging run will be required a year after the discovery observations
simply to recover the phase.

\subsection{Minimizing Potential Contamination to the Transit Signature}
\label{sec:contamination}
\label{sec-candidates}

It is important to select the very best candidates for the RV
follow-up in order to have a high yield of planets.  The three main
characteristics intrinsic to transiting planet light curves are (1)
they show very shallow eclipses, (2) the eclipses have a flat bottom
in a bandpass where limb darkening is negligible, and (3) there is no
secondary eclipse.  Four different types of systems could be
confused with a transiting planet: grazing eclipsing binary stars; an
eclipsing binary system consisting of a large primary star with a small
stellar companion; an eclipsing binary star contaminated by the
light of a third blended star; and a transiting brown dwarf or late M
dwarf.  This section discusses the first three types of possible
contaminants and some ways to differentiate them from bona-fide
planet transit light curves before the RV follow-up.  Contaminants of
the fourth type (brown dwarfs or late M dwarfs) can only be
distinguished by RV follow-up observations, but they are of great
interest in their own right.

\subsubsection{Ruling Out Grazing Eclipsing Binaries}
\label{sec-grazingbinary}

\begin{figure}
\plotone{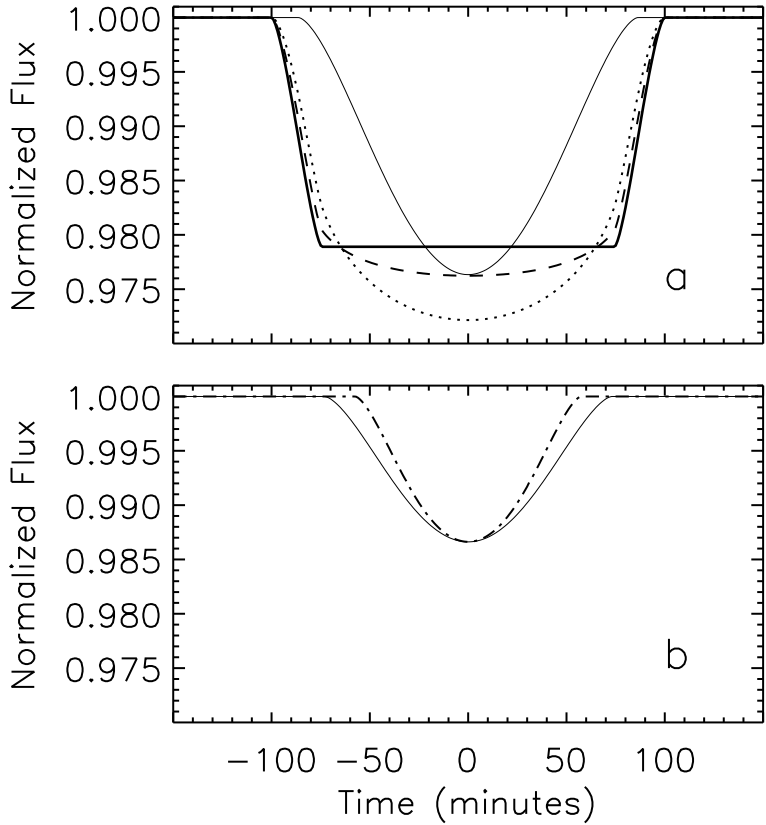}
\caption{A comparison of a planet transit
light curve and a grazing eclipsing binary star light curve.  The
planet has $R=1.4 R_J$ orbiting a star with $R_*=R_{\odot}$, at an
orbital distance of $D=0.05$~AU, with a corresponding period of 4.08
days.  The grazing binary system (thin solid line) is composed of two
identical sun-like stars with twice the planet's period.  Panel a: The
eclipsing binary star system light curve (thin solid line) with solar
limb darkening at $I$, has a grazing angle of $85.11^{\circ}$, chosen
to match a planet transit (at $i=90^{\circ}$) depth in the $I$ band
(dashed line).  Also shown are transit light curves neglecting limb
darkening (thick solid line) and with solar limb darkening at 0.45
$\mu$m (dotted line).  Panel b: The grazing eclipsing binary system
light curve (thin solid line) has a grazing angle 85.024$^{\circ}$
chosen to match the depth of a partial planet transit (dot-dashed
line; the same partial planet transit curve is shown by the dot-dashed
line in Figure~\ref{fig:inc}c).  Panel a illustrates that with high
precision photometry and good time sampling, it should be possible to
distinguish a full planet transit from a grazing binary star eclipse.
Panel b shows that in the case in which a planet produces a partial
transit, the light curve can be nearly identical to that produced by a
grazing binary star.  However, since $R_p  \ll  R_*$ for giant planets
around sun-like stars, partial planet transits are rare.}
\label{fig:graz}
\end{figure}

At certain orbital inclinations a grazing eclipsing binary star can
produce the sought-after drop in brightness of 1\% when a small part
of the companion crosses the primary star.  If the stars are of
similar surface brightness, or if one star has a much larger surface
brightness than the other, then it may not be possible to discern
any secondary eclipses in the data.  Hence, these very shallow
eclipses can be the major cause of false-positive planet candidates in
some transit searches (W.  Borucki 2001, private communication; D.
Latham 2001, private communication).  Even though the eclipse depth
may be the same as a transiting planet, a grazing eclipse from a
binary star system has a different shape.  As illustrated in
Figure~\ref{fig:graz}, a triangular light curve with a rounded bottom
is indicative of a grazing binary system, since the stellar companion
only partially overlaps the primary star's disk.  In contrast, a
transiting planet has a flat-bottomed eclipse, which indicates that
the eclipsing companion is entirely superimposed on the disk of the
primary star.  Note that for a small range of orbital inclinations,
the transiting planet is never fully superimposed on the primary star and
produces an eclipse with a very similar shape and depth to that of a
grazing binary (Figure~\ref{fig:graz}b).  However a partial transit
geometry is rare for $R_p  \ll  R_*$, and in most cases the depth of a
partial transit will be much less than 1\%.  Thus for practical
purposes, even if partial planet transits are not included
in the followup RV measurements they can be accounted for statistically.

High time sampling and high-precision photometry are required in order
to determine the shape of the eclipse, and thus distinguish between
the shallow round eclipses caused by grazing binary stars and
eclipses with flat bottoms that may be caused by transiting planets.
Distinguishing between the two types of eclipses is easiest when the
light curve is taken in a bandpass which is not severely affected by
limb darkening (Figure~\ref{fig:limbdarkening}).

To distinguish grazing binaries from transit candidates, the EXPLORE
project takes observations at a very high rate (every $\lesssim 3$
minutes), and uses an $I$-band filter so that limb darkening is not
significant.  Figure~\ref{fig:grazing}b shows an example of a grazing
binary system light curve from the EXPLORE~I search, as evidenced from
the round bottom and highly sloped ingress and egress of the eclipses.
Although grazing binaries can be trivially ruled out by follow-up RV
measurements, it is essential to have flat-bottomed eclipse candidates
for a high yield of actual planets among the candidates chosen for
follow-up.

\subsubsection{Ruling Out Eclipsing Binary Systems With a Large Primary Star}

A small star eclipsing a large star can have the same eclipse depth as
a Jupiter-sized planet eclipsing a sun-sized star.  For example, an M4
dwarf eclipsing an F0 star will cause a 1\%-deep eclipse with a flat
bottom.  A secondary eclipse is a definite indicator of an eclipsing
binary star system regardless of the eclipse depth.  However, if the
surface brightness ratio of the primary to secondary star is large,
the resulting secondary eclipse will not be visible in the light
curve.  A binary star system with a large primary is easy to rule out
from the length of the eclipse alone.  Figure~\ref{fig:4343}a shows a
clear example of a 2\%-deep eclipse where the Jupiter-size
planet/sun-like star hypothesis can be immediately ruled out.  The
eclipse has a 2.2 day period and lasts 5.5 hours, which is much longer
than an eclipse caused by a planet with the same period orbiting a
solar-type or smaller star. Another way to rule out eclipsing binaries
with a large primary is to consider the unique solution of a light
curve with two or more transits.  With a light curve of sufficient
photometric precision and time sampling, the stellar size and mass can
be derived using the five equations in \S\ref{sec-uniqueness}.  Note
that in the case of a giant star, the eclipse will be much longer than
for a main sequence star of the same mass; solving the five equations
in
\S\ref{sec-uniqueness} using the wrong mass-radius relation
will give a value of $R_*$ which is smaller
than a giant star but still significantly larger than the sun.  This
will likely be enough to rule out the planet hypothesis, and can be
confirmed using the color of the star.  Alternatively, a
binary star system with a large primary can also be ruled out by
spectral classification of the star.

\subsubsection{Ruling Out the Presence of a Contaminating Blended Star}
\label{sec-blends}

A flat-bottomed and relatively deep eclipse from a companion star
fully superimposed on its larger primary will appear shallower if
light from a third blended star is present in the light curve.  The
contaminating blended star could be present due to a chance alignment
with the eclipsing binary system or, more likely if the field is not
too crowded, the contaminating star could be a component of an
unresolved multiple star system.

\begin{figure}
\plotfiddle{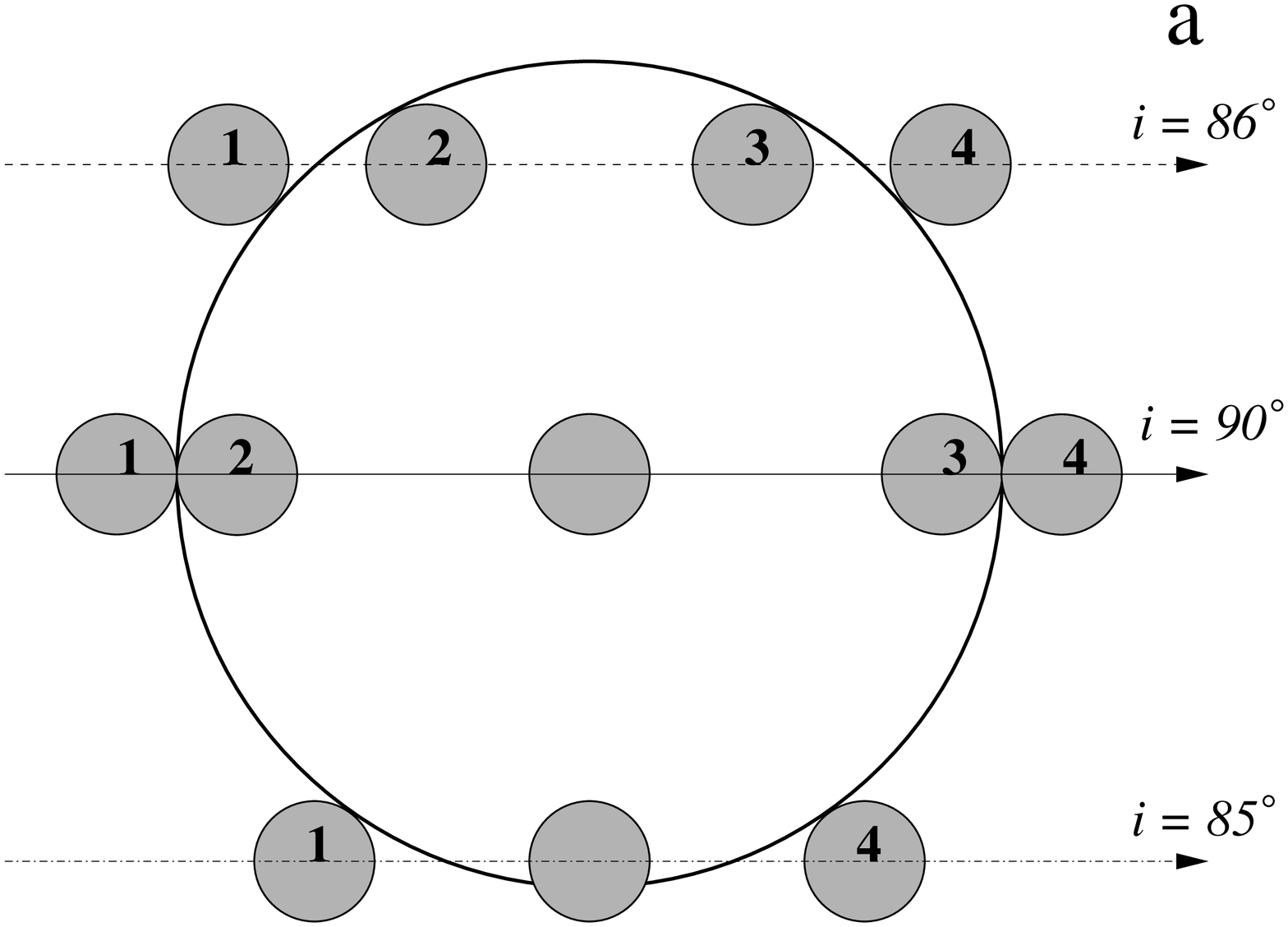}{2in}{0}{35}{35}{-120}{90}
\plotfiddle{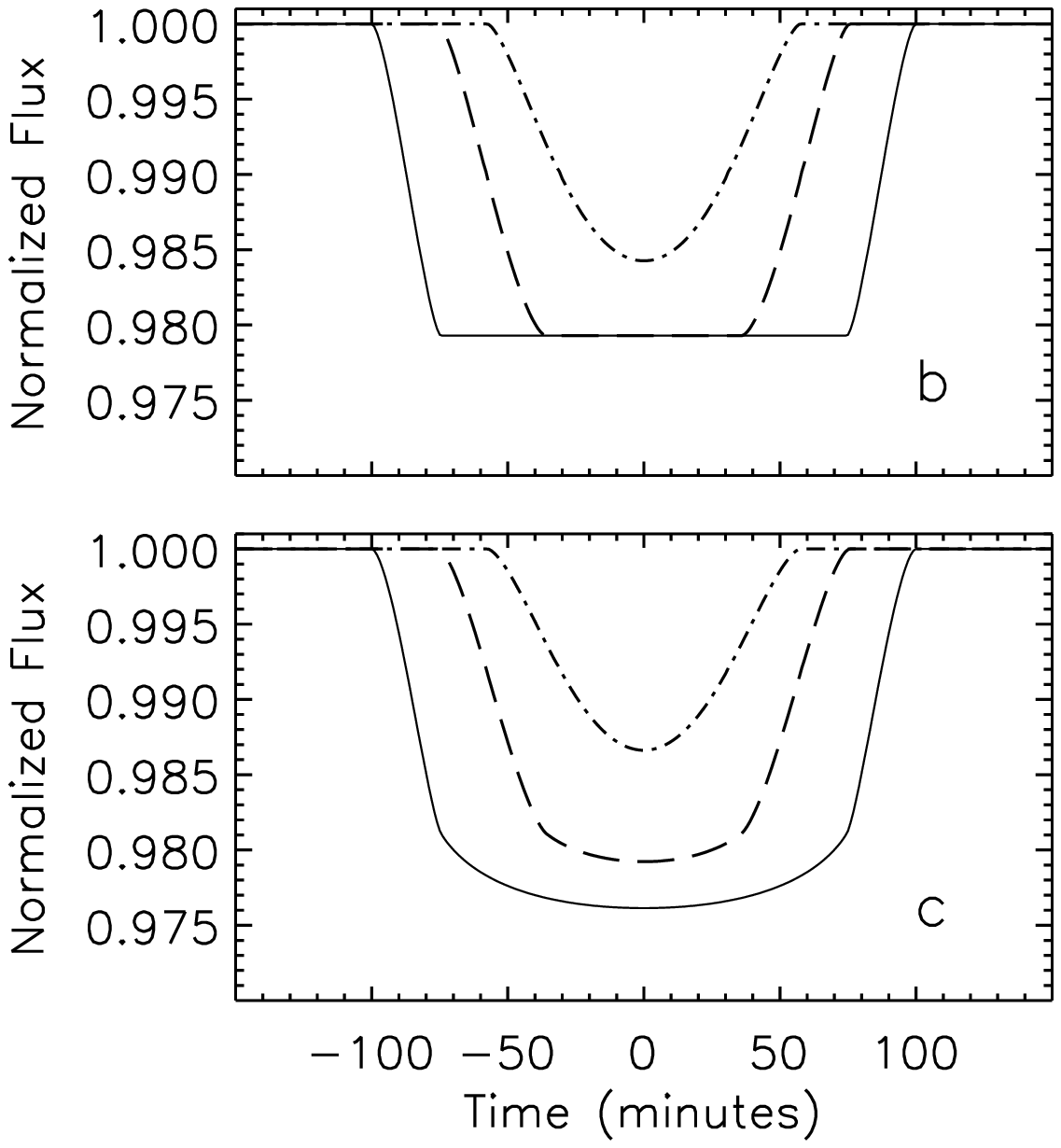}{2in}{0}{75}{65}{-220}{-240}
\caption{
Panel a: A schematic diagram of transiting planets with different
orbital inclinations.  First, second, third, and fourth contact are
indicated.  For lower inclinations the eclipses are shorter and ingress
and egress are longer than for central transits.  The planet and star
are to scale for $R_p=1.4 R_J$ and $R_*=R_{\odot}$.  Panel b: Transit
light curves for the three inclinations shown in panel a, for
parameters $R_p = 1.4 R_J$, $R_*=R_{\odot}$, $D=0.05$~AU.  The orbital
inclinations, from top to bottom are, 85$^{\circ}$, 86$^{\circ}$, and
90$^{\circ}$.  The top curve is round-bottomed because the planet is
only partially transiting the star (see panel a).  At all other
orbital inclinations the transit light curve is flat, indicating that
the planet is fully superimposed on the parent star.  Panel c: The
same transit light curves shown in panel b, but with solar limb
darkening at 0.8$\mu$m adopted.}
\label{fig:inc}
\end{figure}

The unique solution of a light curve with two or more transits can be
used to identify an eclipse contaminated by a blended star.  The
length of the ingress or egress is set by a combination of $R_c/R_*$,
and the projected impact parameter $\frac{D}{R_*} \cos i$ at which the
companion crosses the center of the stellar disk (where $R_c$ is the
radius of the eclipsing companion, $R_*$ is the radius of the central
star, $D$ is the orbital distance, and $i$ is the orbital
inclination).  A 1\% eclipse with an ingress and egress which are long
compared to the total eclipse duration can be produced by the
following two cases: (1) a planet crossing a sun-sized star with
impact parameter $\frac{D}{R_*} \cos i \lesssim 1$ (i.e., the planet
transits near the stellar limb), and (2) a small star eclipsing a
sun-sized star, with additional light from a blended star
contaminating the light curve.  In case (1), the long ingress and
egress are due to the fact that the planet transits close to the limb
and is partially superimposed on the stellar disk for a relatively
long time.  This is illustrated in Figure~\ref{fig:inc}, for the
$i=86^{\circ}$ case (top line in panel a, dashed line in panels b and
c).  Note that only a very small range of inclinations will result in
a transit with a proportionally long ingress/egress; therefore for
bona fide planets, having a transit with a long ingress/egress is much
less likely than one with a short ingress/egress.  In case (2), the
long ingress and egress are due to the fact that a larger companion
will necessarily take a long time to completely cross the primary
star's
limb even for a central eclipse (middle line in
Figure~\ref{fig:inc}a).  Normally cases (1) and (2) would not be
confused because the larger companion in case (2) will produce a much
deeper eclipse than the small companion in case (1).  However, if the
light curve is contaminated with additional light from a bright
blended star, the observed eclipse depth will be reduced, and thus
case (2) can mimic the shallow transit in case (1).  The surface
brightness ratio of the primary and secondary stars in the eclipsing
binary system in case (2) can easily be large enough so that the
secondary eclipse is lost in the photometric noise of the blended
light curve.

The case of an eclipsing binary system plus a blended star can be
ruled out with by using a spectral type to complement the unique
solution to the equations in \S\ref{sec-uniqueness}, provided that the
light curve has two eclipses with good photometric precision and time
sampling.  In the presence of a blended star, the unique solution gives a
stellar mass and radius that are different from the mass and radius
derived from the spectral type.  Specifically, the unique solution
will give an erroneously large primary star.  This is because the
inferred inclination of the orbit will result in a solution in which
the planet transits close to the stellar limb and therefore seems to
be going across a relatively small length of the primary star.  A
spectral type which is inconsistent with the primary star's mass and
radius as derived from the unique solution to the light curve is
therefore a strong indication that there is a contaminating blended
star.  

Even without a spectral type, the planet hypothesis can be ruled out
from the light curve alone in the case of an eclipsing binary system
plus a blended star based on the overestimated primary radius.  The
large stellar radius together with the measured $\sim$
1\% transit depth will usually give a companion radius too large to be
a planet.  In other words, for a given eclipse length, the inferred
stellar radius will be much larger than the true radius, and the
system will appear to be a large star with a smaller stellar companion
transiting very close to the stellar limb.  The uncertainties in the
parameters derived from the unique solution can be large and in some
blended cases the companion can appear to have a size which is almost
compatible with that of a giant planet. Therefore obtaining a spectral
type prior to the RV follow-up is very worthwhile, especially when the
ingress and egress of a transit are long compared to the total transit
duration.

If the data are noisy, it may not be possible to rule out a
contaminating star from the light curve, even if a spectral type is
available.  In this case, radial velocity follow-up observations can
be used to rule out a planet by identifying two components in the
spectrum: (1) a constant-velocity component coming from the blended
star, and (2) a component from the primary star in the eclipsing
binary system that will exhibit large radial velocity changes with the
period and phase corresponding to the transits.  If a bright guide
star or laser guiding is available, it is also possible to test the
blend hypothesis by obtaining a very high resolution image with an
adaptive optics system to look for close companions.  Typical stars in
the EXPLORE search are at 1--2 kpc away, so a resolution of 0.05$''$ would
be adequate to detect companions at 50--100 AU.

\subsubsection{Ruling Out Other Sources of Contamination}

Stellar secular variability might be thought of as a concern in the
search for transits.  However, a transit signal is very different from
most intrinsic variability of a star.  In particular, notice that a
transiting planet causes a drop in brightness during only a small
percentage ($<5\%$) of the total time.  A large spot on the star's
surface, for example, can cause a periodic drop in stellar brightness
with the stellar rotation period.  However, the probability is low for
the combination of the spot's position on the star and the inclination
of rotational axis to conspire to produce a drop in brightness for a
time much shorter than half of the rotational period.  Moreover a star
with one large spot is also likely to exhibit other variability.
Observations at different wavelengths should distinguish variability
due to star spots from a relatively gray planet transit. See Giampapa
et al. (1995) for a thorough discussion of this.  Another concern
might be confusion arising from brown dwarf or M dwarf eclipses;
however, these are of great interest themselves and will be revealed
by follow-up radial-velocity observations.

\section{The EXPLORE~I Transit Search}
\label{sec-light curves}
\label{sec-explore}
We now turn to early results from the first EXPLORE transit search.
The EXPLORE I search was conducted during 11 nights on the CTIO 4-m
telescope with the MOSAIC II camera, on May 30, and June 1-10, 2001.
The MOSAIC II camera is made up of 8 2Kx4K thinned CCDs with 15 $\mu$m
pixels, which corresponds to 0.27$''$/pixel and a $36'\times36'$ field
of view at the CTIO 4-m prime focus.  We observed a single field near
the galactic plane ($l=-27.8$, $b=-2.7$, $\alpha$=16:27:28.00,
$\delta$=--52:52:40.0 J2000.0) with 100,000 stars down to $I=18.2$,
and $\sim350,000$ stars to $I=21$.  At the latitude of CTIO, it was
possible to observe the EXPLORE~I field throughout the whole night,
which resulted in as much as 10.8 hours of coverage on good nights.
A total of $\sim$1800 images were obtained, although $\sim$200 of
these were of very low quality due to poor weather conditions.  The
EXPLORE I field is located in an area of low and uniform dust
extinction, and sits in the Norma patch of the OGLE microlensing
search\footnote{http://bulge.princeton.edu/\~{}ogle/ogle2/fields.html}
(Udalski et al.\ 1997).  It will thus be possible to study long-term
variability of objects in our sample when the OGLE data become public.

\begin{figure}
\plotone{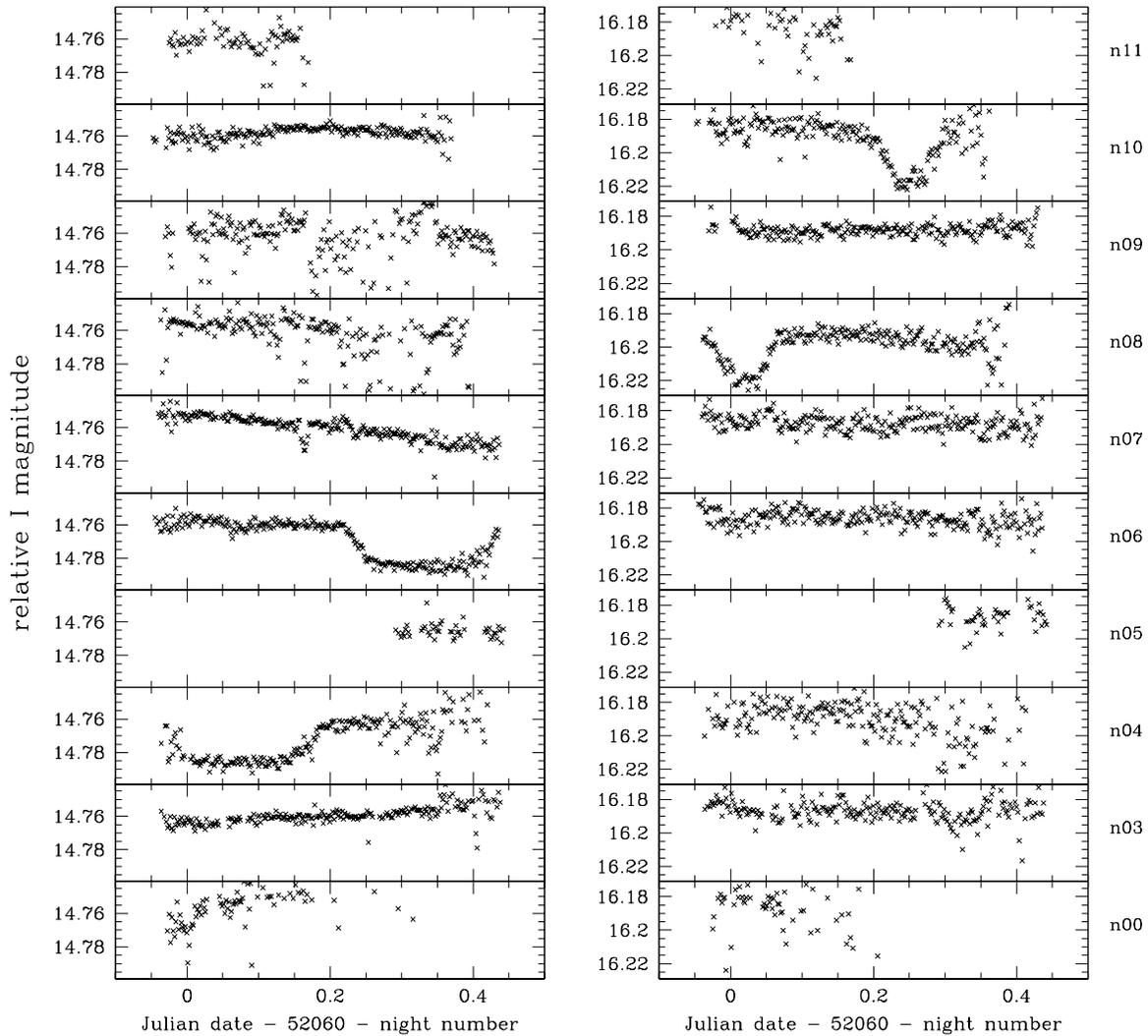}
\caption{Two eclipsing binary star systems from the EXPLORE I search.  
Panel a: The eclipse depth in this light curve is the same depth
($\sim 2$\%) as a giant planet transiting a sun-sized star.  However
the long transit duration ($\sim$ 7\% of the total orbital period)
together with the 2.2 day orbital period indicates that the primary
star has a very large radius, and therefore the companion is too large to
be a planet.  Panel b: A light curve showing two flat-bottomed
eclipses.  The eclipse depth is 3\% and this star appears to have
spectral type early K.  The flat bottom of this light curve means that
the companion is fully superimposed on the primary during the transit.
As discussed in the text, the companion in this system is not a
planet.}
\label{fig:4343}
\end{figure}

\begin{figure}
\plotone{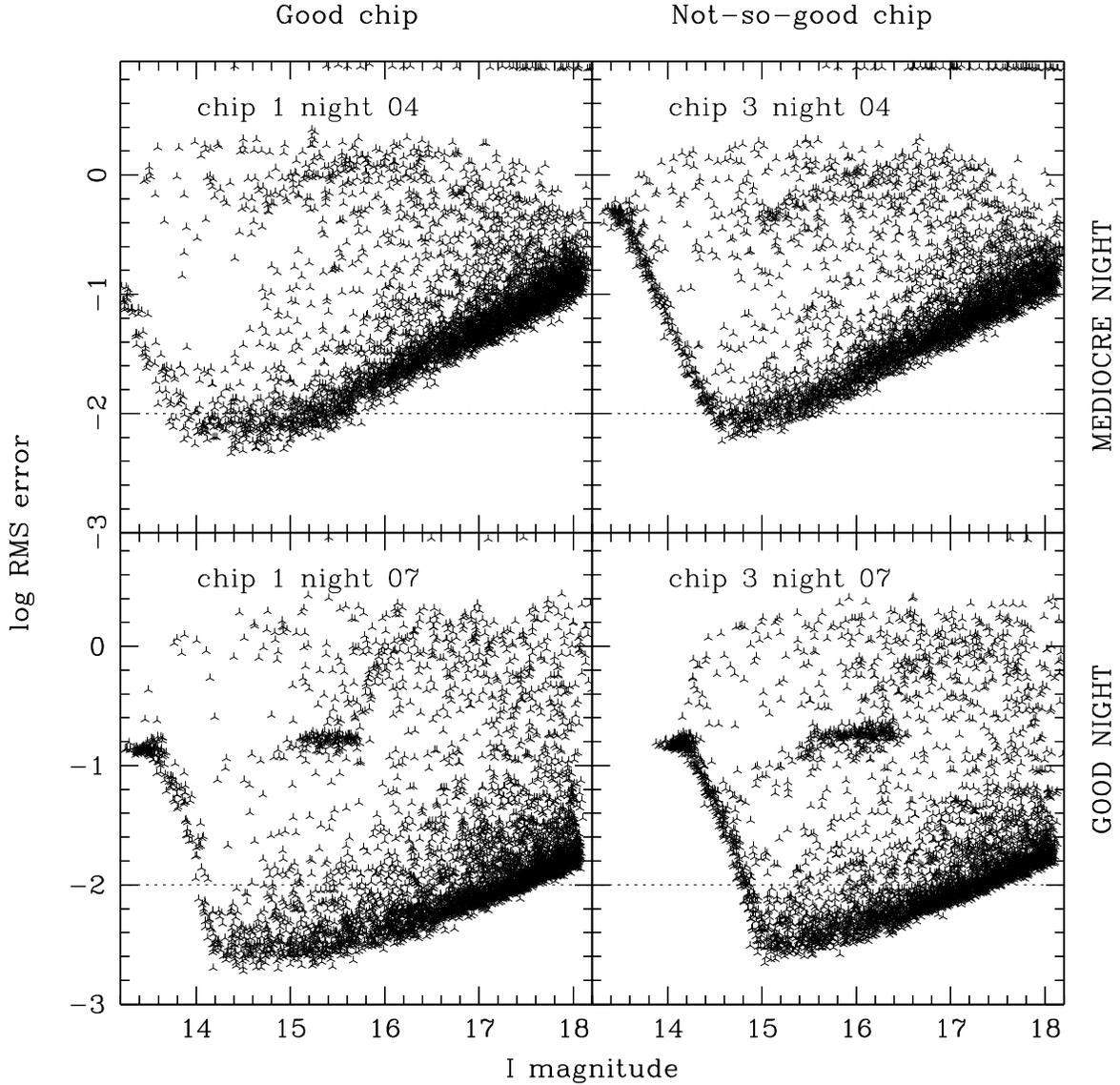}
\caption{Photometric precision from the EXPLORE I search
on the CTIO 4-m MOSAIC camera.  Log rms error vs.\ $I$ magnitude for a
good chip and a not-so-good chip (with shallower electron wells), on a
full good night and a full mediocre night (variable clouds and
variable bad seeing) in our June 2001 run.  Points below the
horizontal dotted line correspond to stars with $<$1\% rms
photometry, which that are suitable for planet detection.  On a bad
night (not shown) there are {\it no} light curves with this precision.
The dense clusters of points at bright magnitudes and the dense
clusters of points with log rms $\sim$ 0.8 are due to saturated
stars.}
\label{fig:photacc}
\end{figure}

\begin{figure}
\plotone{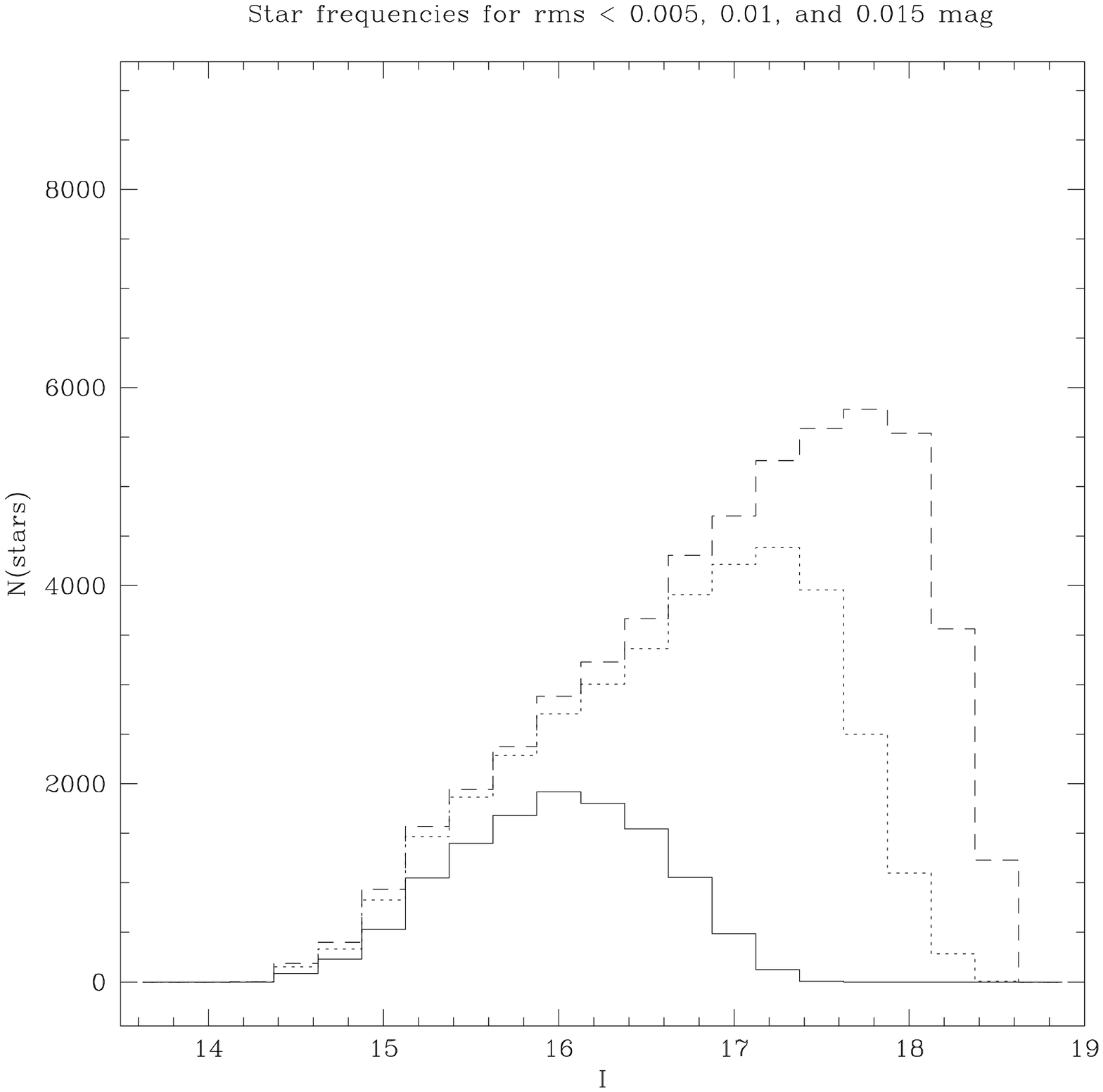}
\caption{Histogram showing the number of stars with rms photometric precision
better than 0.5\% (solid line), 1\% (dotted line), and 1.5\% (dashed
line).}
\label{fig:histogram}
\end{figure}

We had clear weather for approximately 6 nights, and had variable
thick clouds and rain or fog during the rest of the run.  For a
percentage of the stars, however, the data taken during bad weather
still yielded useful photometry.  Our typical exposure time was 60
seconds for good conditions, and was adjusted up to compensate for
cloud cover, or down if the sky was too bright due to moonlit clouds.
The detector read time plus overhead was 101 seconds, so we typically
obtained one photometric data point every 161 seconds, or 2.7 minutes.
Out of the 100,000 stars to $I\sim18.2$ that were initially processed,
we have so far produced 37,000 light curves with 0.2--1\% rms over a
good night (Figure~\ref{fig:photacc}).  
Figure~\ref{fig:histogram} shows a histogram of the number of stars
with photometric precision better than 0.5\%, 1\%, and 1.5\%.  Our
best candidates tend to have $\sim 0.5$\% photometry and $15 \lesssim
I \lesssim 17$.  Figures~\ref{fig:grazing} and \ref{fig:4343} show
sample light curves.

The high photometric precision and high time sampling of our data are
crucial for distinguishing between transits and grazing eclipsing
binary stars (see \S\ref{sec-grazingbinary}), and necessary for
deriving the system parameters from the light curve itself (see
\S\ref{sec-uniqueness}; Seager \& Mall\'en-Ornelas 2002).  Typical
transit lengths for known close-in planets would be 2.5--3 hours
(e.g., HD~209458~b), yielding $\sim 50$ points at minimum light to
0.2--1\% precision (e.g., $7 \sigma$ detections per full transit with
$\delta m = 0.002-0.01$).  Note that a transit of an HD~209458~b-like
planet ($R_p = 1.35 R_J$\footnote{where $R_J$ is Jupiter's equatorial
radius}, Brown et al.\ 2001; see also Cody \& Sasselov 2002) around a
K0 star (typical for our field) will produce an easily detectable
2.6\%-deep transit.

The data were reduced as outlined in \S\ref{sec:photpipe} and
candidates were found by visual examination within seven weeks after
the observing run.  The fast processing and examination of the data
were done so that we could obtain radial velocity follow-up
measurements on our best candidates before phase information was lost.
The final analysis of the data will be done with an automatic
transit-detection program, and extensive tests to determine detection
thresholds and completeness.  A preliminary estimate based on the
window function for the poor weather conditions during our run results
in $\sim$1 expected planet for the 37,000 stars examined so far.  A
statistical analysis and constraints on close-in giant planet
frequency will follow in a later paper.  We followed-up our three most
promising candidates in September 2001 with 19 hours of Director's
Discretionary Time on the VLT+UVES.  Details are discussed in
Mall\'en-Ornelas et al.\ (2002).


We found several systems with eclipses of 1.5\% to 3\% depth.  Three
of these systems each had two flat-bottomed eclipses
(EXP1J1628-52c07s24763, EXP1J1628-52c07s18161, EXP1J1628-52c01s52805).
In addition, we found several eclipsing systems where the noise was
too great to determine the eclipse shape.  Of the three stars showing
two eclipses with flat-bottoms, the first (EXP1J1628-52c07s24763) is
clearly ruled out as a planet candidate because the transit is too
long, implying a large primary (Figure~\ref{fig:4343}a).  The second
system (EXP1J1628-52c07s18161) has a noisy light curve and more data
are needed to confirm the transits.

The third system (EXP1J1628-52c01s52805), shown in
Figure~\ref{fig:4343}b, has two high-quality eclipses with clear flat
bottoms.  Based on follow-up radial velocity data, we now know that
the eclipse is not caused by a planet (Mall\'en-Ornelas et al.\ 2002)
but it is still an interesting system worth discussing.  As is seen in
Figure~\ref{fig:4343}b, the light curve has two 3\%-deep flat-bottomed
eclipses due to a companion in a 2.23-day period orbit, and there is
no sign of a secondary eclipse at $P$/2.  The clear flat bottom
eclipses indicate that the companion disk is fully superimposed on the
parent star.  $BVRI$ photometry and a classification spectrum indicate
that this is an early K star on the main sequence
($M_*\sim0.8~M_{\odot}$ and R$\sim0.85$~\rsun).  A straightforward
interpretation would be that the eclipsing companion is a 1.4~$R_J$
planet orbiting at 8 stellar radii from the parent star.

This third system's eclipse ingress and egress are nearly the same
duration as the flat part duration, which could be due to a planet with an
orbital inclination such that it transits close to the stellar limb
(Figure~\ref{fig:inc}).  However, the almost equal length
ingress/egress and flat part of the light curve is a cause for
concern: applying the uniqueness criteria for a two-transit light
curve (\S\ref{sec-uniqueness}) gives the stellar radius of an F star,
while the spectral type indicates a K star.  As was discussed in
(\S\ref{sec-blends}), this is an indication of the likely presence of
light from a blended companion diluting the light curve and making an
otherwise deep eclipse appear shallow.  Since EXP1J1628-52c01s52805
was our cleanest two-transit light curve, it was included in the
follow-up radial velocity observations.  The observations revealed a
contaminating star with constant RV, blended with a fainter star
showing RV variations with a 2.23 day period and a $\sim$60 km/s
amplitude (see Mall\'en-Ornelas et al.\ 2002).  Thus, for this
eclipsing system, the planet hypothesis was ruled out.

\section{Summary and Conclusion}

The first successful detection of a planet by the transit method will
mark a huge step forward in planet detection and characterization.
Transit searches will allow planets to be found around a variety of
stellar types and in a variety of environments.  Every planet
discovered by the transit method will have a measured radius and
absolute mass (together with radial velocity follow-up measurements)
which will provide key constraints for planetary formation and
evolution models.  In order to find a transiting planet, many
thousands of stars must be monitored for $\gtrsim 2$ weeks with high
photometric precision and high time sampling.  Although many groups
around the world are conducting transit searches and a few groups have
produced transiting planet candidates, no planets have yet been
confirmed by a mass measurement.  In addition to searching for
planets, transit searches are providing databases of variable star
light curves with unprecedented time sampling and photometric
precision.  We note that these databases will also be extremely useful
for finding moving objects such as asteroids, finding short-duration
microlensing events such as those from free-floating planets, and
generating deep images of stellar fields.

The EXPLORE project is a series of transit searches around stars in
the Galactic plane using wide-field large-format CCD cameras on
4-m-class telescopes.  We have presented the EXPLORE transit search
strategy which involves monitoring a single low Galactic latitude
field continuously with high-precision photometry for as many
consecutive nights as possible.  We have shown that continuous
monitoring is the most efficient way to find planet transit candidates
from the ground when only a few weeks of telescope time are
available.  One of the key aspects of our strategy is high time
sampling, which allows us to use the unique solution of a light curve
that has at least two flat-bottomed eclipses in order to select the
very best planet candidates (e.g., by ruling out grazing binary
stars).  A unique aspect of our strategy among current planet transit
searches is that we follow-up our transit candidates with radial
velocity measurements in the same season they were discovered.  This
is key to interpreting the radial velocity observations because only a
few radial velocity points are needed to rule out brown dwarfs and M
dwarfs if the phase is known.  The EXPLORE team has so far
conducted a Southern search with the CTIO 4-m telescope (June
2001) with RV follow-up done on the VLT, as well as a Northern search
with the CFHT 3.6-m telescope (Dec 2001) with RV follow-up done on the
Keck telescope.

We have reported early results of the EXPLORE I search, which used the
CTIO 4-m telescope and MOSAIC~II camera for 11 nights in May/June
2001, out of which $\sim$6 had excellent weather.  We have reached a
photometric precision of 0.2--1\% with points every 2.7 minutes for a
sample of 37,000 stars in the Galactic plane at $l=-27.8$, $b=-2.7$
($\alpha$=16:27:28.00, $\delta$=--52:52:40.0 J2000.0).  For this
number of stars and our limited time coverage, we expect to find $\sim$1
transiting planet.  We have followed up three planet transit
candidates with VLT UVES in September 2001.  The results of the UVES
follow-up are described in Mall\'en-Ornelas et al.\ (2002).


\acknowledgments
We thank the following people for valuable contributions to this
project: George Hau, Sara Ellison, Jon Willis, Laurent Eyer, Fred
Courbin, Beatriz Barbuy, and Mike Wevrick.  We also thank Bohdan
Paczy\'nski, Robert Lupton, Scott Gaudi, and David Charbonneau for
useful conversations.  We thank the Red-Sequence Cluster Survey
project (Howard Yee, Mike Gladders, Felipe Barrientos, and Pat Hall)
for obtaining test field images during their CTIO runs for selecting
the EXPLORE I field.  Andrzej Udalski, Przemyslaw Wo\'zniak, and
Bohdan Paczy\'nski kindly provided us with OGLE color images which was
helpful in selecting the EXPLORE I field.  We thank David Spergel for
letting us use Princeton's Beowulf cluster (Fluffy) which was
essential for the EXPLORE I project.  We would like to thank both the
US and Chile CTIO time allocation committees for generous allocations.
The staff at CTIO were very helpful in accommodating a run which pushed
the telescope data acquisition to its limits.  G.M.O.\ thanks Scott
Tremaine for very valuable discussions, advice, and encouragement.
G.M.O.\ thanks John Bahcall and the IAS for generous support during a
visit when much of this work was carried out.  S.S.\ thanks John
Bahcall for his very strong interest and support for this project.
G.M.O.\ is supported in part by Fundaci\'on Andes and Fondecyt, and
S.S.\ is supported by the W.M.\ Keck Foundation.  The research of H.Y.
and M.G.  is supported in part by NSERC and a grant from the
University of Toronto.  D.M.  is supported by FONDAP Center for
Astrophysics 15010003 and Fondecyt.  The National Center for Atmospheric
Research is supported by the National Science Foundation.

\begin{references}
\reference{} Alard, C., \& Lupton, R.\  1998, ApJ, 503, 325

\reference{} Borucki, W.~J., Caldwell, D., Koch, D.~G., \& Webster,
L.~D.\  2001, PASP, 113, 439

\reference{} Borucki W.~J., \& Summers, A.~L.\  1984, Icarus,
58, 121


\reference{} Brown, T.~M., Charbonneau, D., Gilliland, R.~L., Noyes, R.~W., \& Burrows, A.\  2001, ApJ, 522, 699

\reference{} Brown, T.~M., \& Charbonneau D.
2001, in Planetary Systems in the Universe: Observation, Formation and
Evolution, IAU Symp.\  202, Eds.\  A.\ Penny, P.\ Artymowicz, A.-M.\ Lagrange
and S.\ Russel, ASP Conf.\  Ser.\  in press

\reference{} Burke, C.~J., DePoy, D.~L., Gaudi B.~S., Marshall, J.~L.,
Pogge, R.~W.\ 2002, in {\it Scientific Frontiers in Research on
Extrasolar Planets,} Eds.\ D.\ Demming and S.\ Seager, ASP Conf.\ Ser., in press

\reference{} Butler, R.~P.,  Marcy, G.~W., Williams, E.,  McCarthy, C., Dosanjh, P., \& Vogt, S.~S.\  1996, PASP, 108, 500

\reference{} Butler, R.~P., Marcy, G.~W., Fischer, D.~A., Vogt, S., Tinney, C.~G., Jones, H.~R.~A., Penny, A.~J., \& Apps, K.\  2001, 
in Planetary Systems in the Universe: Observation, Formation and
Evolution, IAU Symp.\  202, Eds.\  A.\ Penny, P.\ Artymowicz, A.-M.\
Lagrange and S.\ Russel, ASP Conf.\ Ser., in press


\reference{} Charbonneau, D., Brown, T.~M., Latham, D.~W., \& Mayor, M.\  2000, ApJ, 529, L45

\reference{} Cody, A.~M.~\& Sasselov, D.~D.\ 2002, ApJ, 569, 451 

\reference{} Dame, T.~M., Hartmann, D., \& Thaddeus, P.\ 2001, 547, 792

\reference{} Gaudi, B.~S.\  2000, ApJ, 539, L59

\reference{} Giampapa, M.~S., Craine, E.~R., \& Hott D.~A.\ 1995,
Icarus, 118, 199
 
\reference{} Gilliland, R.~L.\ et al.\ 2000, ApJ, 545, L47

\reference{} Janes, K.\ 1996, JGR, 101, 14853

\reference{} Henry, G.~W., Marcy, G.~W., Butler, R.~P., \& Vogt, S.~S.\  2000, ApJ, 529, L41

\reference{} Holman, M., Touma, J., \& Tremaine, S.\ 1997, Nature, 386, 254


\reference{} Howell, S.~B., Everett, M.~E., Esquerdo, G., Davis, D.~R.,
Weidenschilling, S., \& van Lew, T.\ 1999, ASP Conf.~Ser.~189:
Precision CCD Photometry, 170


\reference{} Hubbard, W.~B., Burrows, A., \& Lunine, J.~I., 
2002, ARAA, 40, 103

\reference{} Jenkins, J.~M., Caldwell, D.~A., \& Borucki, W.~J.\ 2002, ApJ, 564, 495 

\reference{}  Lin, D.~N.~C., Bodenheimer, P., \& Richardson, D.~C.\  1996, Nature, 380, 606

\reference{} Mall\'en-Ornelas, G., Seager, S., Yee, H.~K.~C., Gladders,
M., Brown, T., Minniti, D., Ellison, S.~E., Mall\'en-Fullerton, G.
2002, in preparation

\reference{} Marcy, G.~W., Cochran, W.~D., \& Mayor, M.\ 2000,
in Protostars and Planets IV, Eds.\  V.~Mannings et al.\ (University of
Arizona Press: Tucson), 1285

\reference{}  Mayor, M.\  \& Queloz, D.\  1995, Nature, 378, 355

\reference{} Mochejska, B.~J., Stanek, K.~Z., 
Sasselov, D.~D., \& Szentgyorgyi, A.~H.\ 2002, AJ, 123, 3460 

\reference{} Monet, D.~B.~A., Canzian, B., Dahn, C., 
Guetter, H., Harris, H., Henden, A., Levine, S., 
Luginbuhl, C., Monet, A.~K.~B., Rhodes, A., Riepe, B., 
Sell, S., Stone, R., Vrba, F., \& Walker, R.\  1998,
The USNO-A2.0 Catalogue, VizieR Online Data Catalog, 1252

\reference{} Murray, N., Hansen, B., Holman, M., \& Tremaine, S.\  1998, Science, 279, 69

\reference{} Pepe, F., Mayor, M., Delabre, B., Kohler, D., Lacroix, D.,
Queloz, D., Udry, S., Benz, W., Bertaux, J.-L., Sivan, J.-P.\  2000,
in  Optical and IR Telescope Instrumentation and Detectors,
Proc., Eds., M.\  Iye and A.~F.\  Moorwood,  SPIE Vol.\  4008, p.\  582

\reference{} Quirrenbach, A., Cooke, J., Mitchell, D., \& Safizadeh, N.\  2001, in Planetary Systems in the Universe: Observation,
Formation and Evolution, IAU Symp.\  202, Eds.\  A.\  Penny, P.\  Artymowicz,
A.-M.\  Lagrange and S.\  Russel, ASP Conf.\  Ser.,  in press

\reference{} Rasio, F.~A., \& Ford, E.\  1996, Science, 274, 954

\reference{} Rosenblatt, F.\  1971, Icarus, 14, 71

\reference{} Sackett, P.\  1999, in Planets Outside the Solar System:
Theory and Observations, NATO ASI, Eds.\  J.-M.\  Mariotti and D.\  Alloin,
(Kluwer: Dordrecht), p.\  189

\reference{} Schlegel, D.~J., Finkbeiner, D.~P., \& Davis, M.\  1998, ApJ, 500, 525

\reference{} Seager, S., \& Mall\'en-Ornelas, G.\ 2002, submitted to ApJ, astro-ph/0206228




\reference{} Street, R.~A., Horne, K., Penny, A., Tsapras, Y.,
Quirrenbach, A., Safizadeh, N., Cooke, J., Mitchell, D., \& Cameron,
A.~C.\  2001, in Planetary Systems in the Universe: Observation,
Formation and Evolution, IAU Symp.\  202, Eds.\  A.\  Penny, P.\  Artymowicz,
A.-M.\  Lagrange and S.\  Russel, ASP Conf.\  Ser.,  in press

\reference{} Street, R.~A., Horne, K., Cameron, A.~C., Tsapras, Y.,
Bramich, D., Penny, A., Quirrenbach, A., Safizadeh, N., Mitchell, D.,
\& Cooke, J.\  2002, in {\it Scientific Frontiers in Research on
Extrasolar Planets, ASP Conf.\ Ser.,} Eds.\ D.\ Demming and S.\ Seager,
in press

\reference{} Struve, O.\  1952, The Observatory, 72, 199

\reference{} Udalski, A., Kubiak, M., \& Szymanski, M.\  1997, Acta Astron, 47, 319

\reference{} Udalski, A.\  et al.\ 2002a, Acta Astronomica, 52, 1 

\reference{} Udalski, A., \.Zebru\'n, K., Szyma\'nski, M., Kubiak, M.,
Soszy\'nski, I., Szewczyk, O., Wyrzykowski, L., \& Pietrzy\'nski, G.\
2002b, Acta Astronomica, 52, 115

\reference{} Yee, H.~K.~C., Mall\'en-Ornelas, G., Seager, S., Gladders, M., \& Mall\'en-Fullerton, G.\ 2002, in preparation

\reference{} Yee, H.~K.~C.\  1991, PASP, 103, 396

\reference{} Yee, H.~K.~C.\  1988, AJ, 95, 1331

\reference{} Wo\'zniak, P.~R.\  2000, Acta Astron., 49, 223

\end{references}
\end{document}